\documentclass{article}
\usepackage{amsmath,amsfonts,amssymb,amsthm,amsopn,amscd}
\usepackage{times}
\usepackage{graphicx}
\usepackage{color}
\usepackage{colortbl}
\usepackage{multirow}
\usepackage{subfigure}
\usepackage{authblk}
\usepackage{ucs} 
\usepackage[utf8x]{inputenc}
\usepackage[normalem]{ulem}
\usepackage[authoryear]{natbib}
\usepackage{rotating}
\usepackage{bbm}
\usepackage{latexsym}

\newcommand{\captionfonts}{\normalsize}

\makeatletter  
\long\def\@makecaption#1#2{%
  \vskip\abovecaptionskip
  \sbox\@tempboxa{{\captionfonts #1: #2}}%
  \ifdim \wd\@tempboxa >\hsize
    {\captionfonts #1: #2\par}
  \else
    \hbox to\hsize{\hfil\box\@tempboxa\hfil}%
  \fi
  \vskip\belowcaptionskip}
\makeatother   

\renewcommand{\cite}{\citep}

\graphicspath{{../Submission/Figures/}}

\newcommand{\der}[2]{\frac{\text{d} #1}{\text{d} #2}}	        

\theoremstyle{plain}
\newtheorem{theorem}{Theorem}[section]

\theoremstyle{definition}

\newtheorem{proposition}[theorem]{Proposition}

\newtheorem{definition}{Definition}[section]

\newtheorem{remark}{Remark}

\theoremstyle{remark}

\newcommand{\R}{\mathbb{R}}
\newcommand{\N}{\mathbb{N}}

\newcommand{\V}{\mathcal{V}}

\newcommand{\Lop}{\mathcal{L}}

\newcommand{\cProb}[2]{\mathbbm{P}\Big [ #1 \big \vert #2 \Big ]}	

\newcommand{\eqdef}{\overset{\rm def}{=}}

\newcommand{\argmin}{\mathop{\mathrm{Arg\,Min}}}

\newcommand{\bd}{\begin{document}}
\newcommand{\ed}{\end{document}}
\newcommand{\beq}{\begin{equation}}
\newcommand{\eeq}{\end{equation}}
\newcommand{\nid}{\noindent}
\newcommand{\ben}{\begin{enumerate}}
\newcommand{\een}{\end{enumerate}}
\newcommand{\bit}{\begin{itemize}}
\newcommand{\eit}{\end{itemize}}
\newcommand{\baR}{\begin{array}}
\newcommand{\eaR}{\end{array}}
\newcommand{\su}{\section}
\newcommand{\ssu}{\subsection}
\newcommand{\sssu}{\subsubsection}
\newcommand{\deq}{\stackrel {\rm def}{=}}
\newcommand{\bea}{\begin{eqnarray}}
\newcommand{\eea}{\end{eqnarray}}

\newcommand {\VV} {\mathbf{V}}
\newcommand{\Diag}{{\mathbf{\rm Diag}}}
\newcommand{\Jb}{\bar{J}}
\newcommand{\Jbab}{\bar{J}_{\alpha\beta}}
\newcommand{\Jbcd}{\bar{J}_{\gamma\delta}}
\newcommand{\Jab}{\sigma_{\alpha\beta}}
\newcommand{\Jdab}{\sigma^2_{\alpha\beta}}
\newcommand{\ma}{m_\alpha}
\newcommand{\Ma}{M_\alpha}
\newcommand{\mua}{\mu_\alpha}
\newcommand{\muax}{\mua^X}
\newcommand{\muay}{\mua^Y}
\newcommand{\va}{v_\alpha}
\newcommand{\vax}{v_\alpha^X}
\newcommand{\Ca}{C_{\alpha\alpha}}
\newcommand{\Cax}{C_{\alpha\alpha}^X}
\newcommand{\Cbx}{C_{\beta\beta}^X}
\newcommand{\Cay}{C_{\alpha\alpha}^Y}
\newcommand{\Ia}{I_\alpha}
\newcommand{\Da}{\Delta_\alpha}
\newcommand{\Deab}{\Delta_{\alpha\beta}}
\newcommand{\Deabx}{\Delta_{\alpha\beta}^X}
\newcommand{\sa}{s_\alpha}
\newcommand{\ta}{\tau_\alpha}
\newcommand{\mub}{\mu_\beta}
\newcommand{\mb}{m_{\alpha\beta}}
\newcommand{\vb}{v_\beta}
\newcommand{\vbx}{v_\beta^X}
\newcommand{\Cb}{C_{\beta\beta}}
\newcommand{\Db}{\Delta_\beta}
\newcommand{\Mb}{M_\beta}
\newcommand{\qb}{q_\beta}
\newcommand{\gb}{\gamma_\beta(\mub,\vb;\Cb)}
\newcommand{\gbt}{\gamma_\beta(\mub,\vb;\Cb(t))}
\newcommand{\tb}{\tau_\beta}
\newcommand{\Sb}{S_\beta}
\newcommand{\Ha}{H_{\alpha}}
\newcommand{\Hax}{H_{\alpha}^X}
\newcommand{\Hab}{H_{\alpha\beta}}
\newcommand{\Habx}{H_{\alpha\beta}^X}
\newcommand{\Dab}{D_{\alpha\beta}}
\newcommand{\Dabx}{D_{\alpha\beta}^X}
\newcommand{\cJ}{{\cal J}}

\newcommand\bbbr{{\sf I\!R}}

\newcommand{\eps}{\varepsilon}

\newcommand{\Cov}{\mathrm{Cov}}

\newcommand{\m}{\mathcal}
\newcommand{\Ie}{I_{\text{ext}}}

\newcommand{\mC}{\mathbf{C}}
\newcommand{\mD}{\mathbf{D}}
\newcommand{\mE}{\mathbf{E}}
\newcommand{\mX}{\mathbf{X}}
\newcommand{\mx}{\mathbf{x}}
\newcommand{\mY}{\mathbf{Y}}
\newcommand{\mL}{\mathbf{L}}
\newcommand{\mQ}{\mathbf{Q}}
\newcommand{\mI}{\mathbf{I}}
\newcommand{\mJ}{\mathbf{J}}
\newcommand{\mW}{\mathbf{W}}
\newcommand{\mU}{\mathbf{U}}
\newcommand{\mone}{\mathbf{1}}

\newcommand{\msig}{\boldsymbol{\sigma}}
\newcommand{\mtheta}{\boldsymbol{\theta}}
\newcommand{\mdelta}{\boldsymbol{\Delta}}
\newcommand{\mla}{\boldsymbol{\Lambda}}
\newcommand{\mga}{\boldsymbol{\Gamma}}
\newcommand{\mxi}{\boldsymbol{\xi}}

\newcommand{\X}{\mathcal{X}}
\newcommand{\proc}{\mathcal{M}_1^+(C([t_0,T], \R)}
\newcommand{\procP}[1]{\mathcal{M}_1^+(C([t_0,T], \R^{#1}))}
\newcommand{\proci}{\mathcal{M}_1^+(C((-\infty,T], \R)}
\newcommand{\prociP}[1]{\mathcal{M}_1^+(C((-\infty,T], \R^{#1})}
\newcommand{\procs}[1]{\mathcal{M}_1^+(C([0,#1], \R)}
\newcommand{\ExpSimple}{\mathbb{E}}
\newcommand{\nvar}[1]{\left \| #1 \right\|_{\mathrm{var}}}
\newcommand{\nvarP}[2]{\left \| #1 \right\|_{\mathrm{var}, #2}}
\newcommand{\abs}[1]{\left \vert #1 \right \vert }
\newcommand{\pente}{\|S_{\beta}'\|_{\infty}}
\newcommand{\Fs}{F_{\text{stat}}}
\newcommand{\triplebar}[1]{\vert\vert\vert #1 \vert\vert\vert}
\newcommand{\procsP}[2]{\mathcal{M}_1^+(C([t_0,#1], \R^{#2})}
\newcommand{\dvar}[2]{d_{\rm{var}}\left(#1,\; #2\right)}
\newcommand{\dt}[2]{d_t\left(#1,\; #2\right)}
\newcommand{\norm}[1]{\left \| #1 \right \|}
\newcommand{\Si}{S}
\newcommand{\IT}{\mathcal{IT}}
\newcommand{\tV}{\mathbf{\widetilde{V}}}

\newcommand{\mA}{\mathbf{A}}

\newcommand{\MyFigure}[5]{
\begin{figure}
 \begin{center}
  \includegraphics[width=#1\textwidth]{#2}
 \end{center}
 \caption[#3]{#4}
 \label{#5}
\end{figure}
}

\newenvironment{heuriproof}{\noindent \emph{Heuristic proof.}}{\begin{flushright} \qed \end{flushright}}

\usepackage{ifthen}
\newboolean{DisplaySOS}
\setboolean{DisplaySOS}{true} 
\newcommand{\SOS}[1]{\ifthenelse{\boolean{DisplaySOS}}{{[#1]}}{}}

\definecolor{violet}{rgb}{1.00,0.00,1.00}	

\newcommand{\john}[1]{{#1}}  
\newcommand{\jsout}[1]{}		
\newcommand{\olivier}[1]{{#1}}
\newcommand{\review}[1]{{{ #1}}}
\newcommand{\reviewo}[1]{ {{ #1}}}
\newcommand{\reviewJ}[1]{ {{ #1}}}
\newcommand{\reviewNew}[1]{{{ #1}}}

\begin{document}

\title{A Markovian event-based framework for stochastic spiking neural networks}
\author[1]{Jonathan Touboul}
\author[1]{Olivier Faugeras}
\affil[1]{\small NeuroMathComp Laboratory, INRIA, Sophia Antipolis, France}

\maketitle


\section*{Abstract}
In spiking neural networks, the information is conveyed by the spike times, that depend on the intrinsic dynamics of each neuron, the input they receive and on the connections between neurons. In this article we study the Markovian nature of the sequence of spike times in stochastic neural networks, and in particular the ability to deduce from a spike train the next spike time, and therefore produce a description of the network activity only based on the spike times regardless of the membrane potential process. 

To study this question in a rigorous manner, we introduce and study an event-based description of networks of noisy integrate-and-fire neurons, i.e. that is based on the computation of the spike times. We show that the firing times of the neurons in the networks constitute a Markov chain, whose transition probability is related to the probability distribution of the interspike interval of the neurons in the network. In the cases where the Markovian model can be developed, the transition probability is explicitly derived in such classical cases \review{of neural networks as } the linear integrate-and-fire neuron models with excitatory and inhibitory interactions, for different types of synapses, possibly featuring  noisy synaptic integration, transmission delays and absolute and relative refractory period. This covers most of the cases that have been investigated in the event-based description of spiking deterministic neural networks.

\section*{Introduction}
Growing experimental evidence tends to establish that spike timings are essential to account for neural computations 
(see e.g.  \cite{delorme-thorpe:01,fabre-thorpe-richard-etal:98,thorpe-delorme-etal:01}). This fact has motivated the use of spiking neuron models, rather than the traditional rate-based models, and a very general interest for the characterization of the structure of spike times as well as for their precise and efficient computation. 

One of the great challenges in \review{computational} neuroscience consists in inferring from a sequence of spike times the next spike that will be fired in the network. In other words, knowing the input a neuron receives and the network's connectivity map, can one compute the dynamics of the spike times regardless of the actual value of the underlying membrane potential value? Answering this question amounts defining what is called an event-based description of the network (see e.g. the review of \cite{brette-rudolph-etal:07}). The purpose of these models is to provide an exact simulation of the network, and this is precisely based on the ability to infer the next spike time based on the knowledge of the previous spike times and on the parameters of the models, without referring to the underlying membrane potential.  

As discussed in \cite{brette-rudolph-etal:07}, in the deterministic case, this property of event-based models is true only for very few cases of simple types of neuron models, where explicit solutions to the membrane potential voltage equation are available. Deterministic event-based algorithms were only developed for simple pulse-coupled integrate-and-fire models \cite{watts:94,claverol-etal:02,delorme-thorpe:03}, and more complex ones in the case of the Spike Response Model \cite{makino:03,marian-etal:02,gerstner-kistler:02}. Rudolph and Destexhe developed a class of models that allow event-based simulation \cite{rudolph-destexhe:06} including a model of synaptic conductances. Romain Brette described in \cite{brette:06,brette:07} algorithms that apply to some classes of input current (exponential currents), and Tonnelier and collaborators extended these works to the case of the quadratic integrate-and-fire models in a case where the equation is integrable in closed-form \cite{tonnelier-etal:07}. Even in this case the authors must make quite restrictive \review{assumptions} with regards to the connectivity and the input current while being more general in terms of the intrinsic dynamics. Their approach relies on the fact that the solution of the particular equation they choose has a closed-form  solution. The price to pay is that no generalization is possible. All these models share the property that the solution to the membrane-potential equation can be found in closed-form, and therefore the spike times can be expressed as the times of the first intersection of these solutions to the threshold, i.e. by inverting explicitly the solution. Moreover, these models only apply to deterministic inputs of a very specific form (either constant, or some types of exponential currents for the linear IF neuron).

They also have the common property that they do not take into account the natural randomness present in the input to the neurons. In the cortex, noise constitutes a prominent aspect of the neural activity, and is even sometimes assumed to have a functional \review{role} (see e.g. \cite{rolls-deco:10}). A common model for this random arrival of spikes consists in considering that the neuron receives a Brownian noisy current (diffusion approximation), as we further explain in section \ref{section:theoretical}. Taking into account the effect of noise in the spike timings is a very difficult task, even for single neurons, and has been extensively studied in the past decades and still constitutes an active area of research (see e.g. the books \cite{holden:76}, \cite{ricciardi:77} and \cite{gerstner-kistler:02} or \cite{touboul-faugeras:07b} and the references therein for a review).

In the stochastic case, event-based descriptions of spike timings become much more complicated. Indeed, even when we have a closed-form expression of the membrane potential (which only occurs in very few specific cases) there are at least two reasons why the problem is extremely difficult to solve. First, it is not always possible to invert the probability density function of the membrane potential to find the first crossing time to the threshold. Second is the fact that the spike times of all neurons can have a very complex correlation structure with the underlying membrane potential voltage: even if we were able to compute the probability distribution of the next spike numerically, the \review{dependence} on the membrane potential variable makes the next spike time extremely difficult to tame.

Extending these results to the statistics of spike times in a network constitutes therefore the next step in both the theory of event-based descriptions of networks and of spike statistics. Indeed, the approach proposed in the deterministic case of solving in closed-form the membrane-potential equation and inverting it will fail. This problem was already identified in the review by Brette and collaborators~\cite{brette-rudolph-etal:07}, where the authors mention that ``the problem of simulating directly a stochastic process in asynchronous algorithms is much harder because even for the simplest stochastic neuron models, there is no closed analytical formula for the distribution of the time to the next spike (see e.g. \cite{tuckwell:88})''. Another extremely complex problem lies in the \review{dependence} structure of the spike times among the different neurons of the network, and the statistical correlation of the next spike time to the trajectory (sample path) of the membrane potential. These are precisely the problems we address in the present manuscript.

\review{Existing work on event-based descriptions of neuronal networks is based on a perfect match between spike times of the voltage process and the event-based model. If this aim is highly relevant in the deterministic case, in the stochastic case the relevant information is the probability distribution of the spike times (i.e. the probability of a given neuron to fire at a given time) rather than some specific realizations of spike trains. Our approach therefore aims at perfectly reproducing the probability distribution of the spike times of the network.}

In all the paper, we consider networks composed of $N$ interacting integrate-and-fire spiking neurons \review{subjected to} noisy inputs from the external environment.
%
%
Mathematically, the membrane potential of these neurons is defined by a stochastic process: each neuron receives at its synapses noisy inputs corresponding to the random activity of ion channels and to the external activity of the network. During the time intervals where no spike is emitted in the network, the membrane potential of each neuron evolves independently according to its intrinsic dynamics. When the membrane potential $V^{(i)}(t)$ of the neuron indexed by $i$ reaches its deterministic threshold function $\theta(t)$ at time $t_0$, the neuron elicits an action potential. Subsequently, its membrane potential is reset to a given value $V_r^{(i)}$, and the states of all its postsynaptic neurons $j$ are modified. We denote by $\V(i)$ the postsynaptic neighborhood of the neuron $i$, i.e. the set of neurons that receive the spikes fired by neuron $i$. When a spike is transmitted to a neighboring cell, it results in the emergence of a positive or negative postsynaptic pulse whose sign depends on the connection type and whose shape depends on the neurotransmitters and  the receptors. 

%
%

%


In this article, we address the problem of event-based models for stochastic networks, and more fundamentally of the Markovian nature of the spike times, in some simple types of networks of integrate-and-fire neurons with different intrinsic dynamics and different kinds of synaptic integration. \review{To this purpose, we introduce a special mathematical framework to describe the dynamics of the spikes in section~\ref{section:theoretical}. We then start by addressing the case of purely inhibitory networks, and we show in section~\ref{section:inhib} that the spike times can be described through the introduction of a Markov chain whose transition probability is related to a first passage-time problem, for some simple types of neuron \review{models}. We further extend this approach in appendix~\ref{append:FurtherModels} to more complex bio-inspired neuronal descriptions, and show the Markov property does not necessarily hold true in all cases. The more slightly complex case of networks including excitatory interactions is then treated in section~\ref{sec:Excitation}, and the same conclusion holds true: a Markov chain can be defined in relation with the spike times in the network, whose transitions are related to first passage time problems. The present framework and derivations open the way to different theoretical, numerical and computational developments, some of which are discussed in the conclusion.}

\section{Theoretical framework}
\label{section:theoretical}
We start by introducing the general mathematical concepts and the framework that we will be using in this manuscript in order to study spike times in stochastic integrate-and-fire networks. \review{The first and central concept we want to recall is the Markov property for a random sequence. Basically, assume that the random sequence $(X_k)_{k\in \N}$ is defined by a certain recurrence property given by the transition probability density $p(X_{n+1}\vert X_{n}, \,\ldots,\, X_0)$. This expresses probability density of the process $X_{n+1}$ conditionally to the values of $X_0,\ldots,X_{n}$. The random sequence is called a Markov chain if this conditional probability depends only on $X_{n}$, being independent of $(X_k)_{k=0,\ldots n-1}$, that is:
\[p(X_{n+1}=x_{n+1}\vert X_{n}=x_{n},\,\ldots,\,X_0=x_0)=p(X_{n+1}=x_{n+1}\vert X_{n}=x_{n}).\] 
Loosely speaking, the knowledge of the present state makes the future of the process independent of its past states. }

The building block of our framework consists in considering the dynamics of the spike times instead of the dynamics of all variables involved in the description of the neuron's state. This purpose motivates the introduction of what we call the \emph{countdown process}, a stochastic process governing the dynamics of the spikes in the network and defined as follows:
\begin{definition} {\it [Countdown process]}
{For each neuron $i$, we define $X^{(i)}(t) \geq 0$ to be the remaining time until the next emission of a spike by neuron $i$ if it does not receive any spike meanwhile. We call this stochastic process the {\it countdown process} of the neuron $i$. \reviewo{Following the tradition established in the stochastic calculus literature, we also sometimes note $X_t^{(i)}$ for $X^{(i)}(t)$.} \review{The $N$ dimensional process $X_t=(X^{(1)}_t, \ldots, X^{(N)}_t)$ is called the countdown process of the network, or simply countdown process}. }
\end{definition}

\review{For each neuron $i\in\{1,\ldots,N\}$, the dynamics of $(X^{(i)}(t))$ is very simple, linearly decreasing with slope $-1$ during the intervals of time where no spike is received or produced: 
\begin{equation}\label{eq:Xdyn}
\frac{dX^{(i)}(t)}{dt} = - 1
\end{equation}
At time $t$, the next spike will be fired by neuron  $i=\argmin_{j\in 1\ldots N} X^{(j)}(t)$ at time $t+X^{(i)}(t)$.}


{At spike time, the membrane potential of the neuron that just fired an action potential is reset, and therefore the new countdown value after the spike emission is updated to a new value corresponding to the next spike time of this neuron if nothing occurs meanwhile. This random variable, noted $Y^{(i)}$, is referred to as the \emph{reset random variable}. 

 Each neuron connected to the one that fired the action potential has his state modified, and \review{its} next firing time is reset accordingly, conditionally to the value of the next firing time predicted before the reception of the spike. }
%
%

%
{The next spike time of neuron $j$ is modified consequently to the reception of the spike from neuron $i$ and the new countdown value of neuron $j$ after spike emission is reset according to a random variable denoted $\eta_{ij}$, corresponds to the next firing time of neuron $j$ after taking into account the reception of the incoming spike from $i$, and conditionally on the countdown value at the spike reception (see Fig \ref{fig:MembraneVsCountdown})\footnote{and possibly on different other variables depending on the model considered.}. This random variable is called the \emph{interaction random variable}.}
\MyFigure{.7}{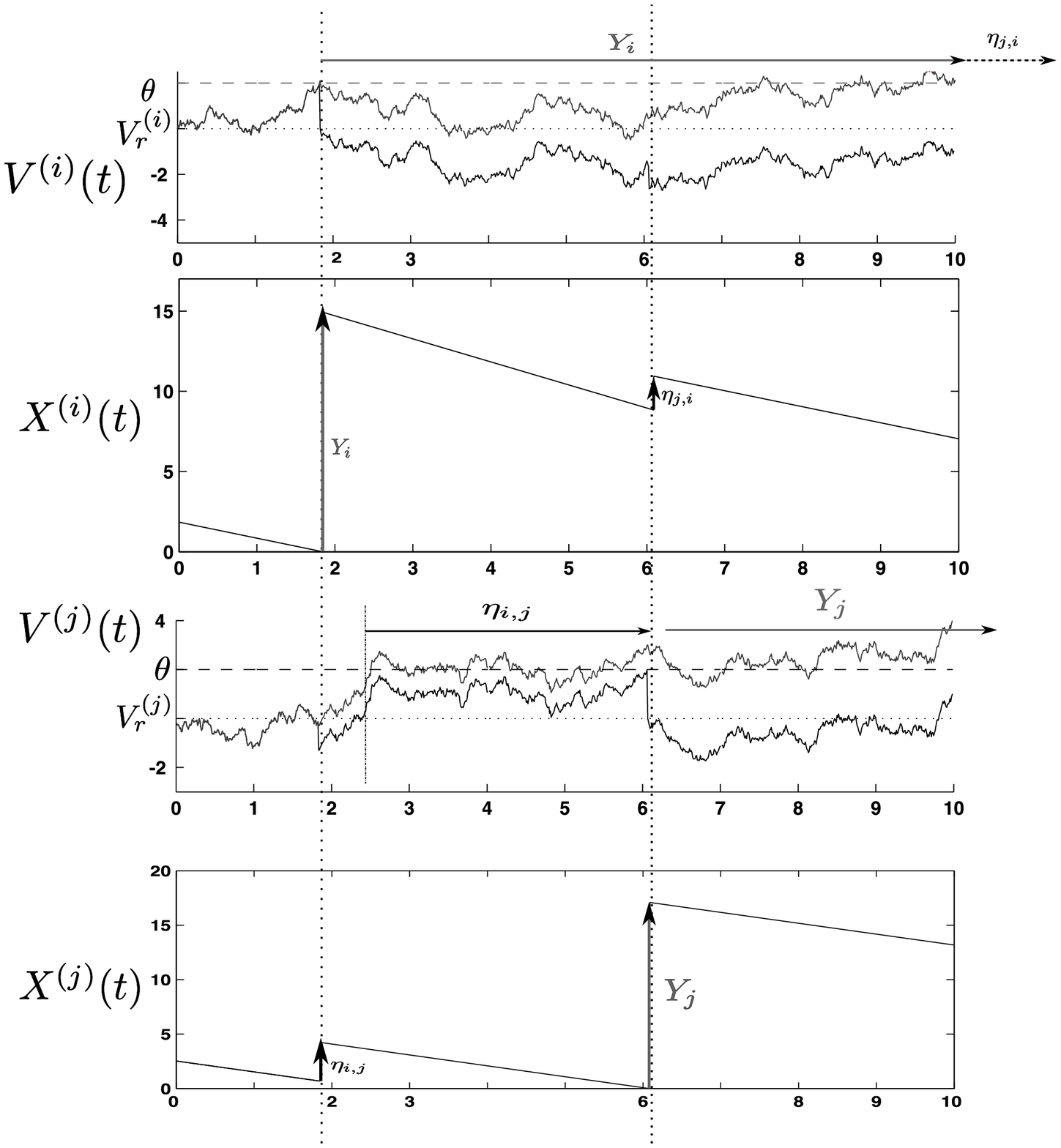}{Countdown Process}{A representation of a sample path for the countdown process and the related membrane potential (case of the perfect integrate-and-fire neurons network).}{fig:MembraneVsCountdown}
{We observe that the countdown process is a piecewise deterministic process such as the ones described in \cite{davis:84}. The spike times $(t_1,\ldots,t_n,\ldots)$ are the only possible discontinuities of the countdown process. The \emph{countdown sequence} is defined as the value of the \review{$N$-dimensional} countdown process \review{of the network} sampled just after the spike emission: $X_n:=X(t_{n}^+)$. It is easy to show that the knowledge of the countdown process is equivalent to the knowledge of the countdown sequence, as we now prove.}

\begin{proposition}\label{prop:EquivChainProcess}
	{There is a path-wise bi-univocal correspondence between the continuous-time countdown process and the discrete-time countdown sequence.}
\end{proposition}

\begin{proof}
	{Assume that we have a realization of the countdown process $(X_t)_{t \in [0,T]}$. The spike times $(t_1,\ldots,t_n,\ldots)$ are given by the zeros of the process, and the value of the process just after these spike times $X_n:=X(t_{n}^+)$ provides the countdown sequence in a unique way.}
	
	{Conversely, assume that we dispose of the values of the countdown sequence $(X_n)_{n\in \mathbb{N}}$. Then we build the related countdown process $\tilde{X}_t$ by ``filling the gaps'' using the deterministic dynamics between two spikes: $\tilde{X}_0=X_0$, and between two spikes, each components $\tilde{X}_t$ decreases linearly with slope $-1$. At time $t_1$ equal to the minimal value of $X_0$, the first spike is fired, and the process $\tilde{X}$ is reset to the value: $\tilde{X}(t_1)=X_1$. The same procedure is then iterated and yields the unique countdown process related to the countdown sequence considered.}
\end{proof}

{This kind of processes has already been studied in the field of queueing theory under the name of \emph{the hourglass model}. It was first introduced by Marie Cottrell in~\cite{cottrell:92}, studied by Fricker, Robert and collaborators in \cite{fricker-robert-etal:94} and then instantiated and studied in ~\cite{asmussen-turova:98, cottrell-turova:00, turova:00, turova:96}. In order to apply this modeling to the kind of networks of interest here, we need to define in each case the reset and the interactions random variables. }

{The framework proposed here is on the equivalence of the spike times in the original network and the spike times emulated by the countdown process (or the countdown sequence). This property actually stems from the definitions of the reset and interaction random variables. We however provide a formal proof since this equivalence is a cornerstone of the paper.}

\begin{theorem}\label{thm:FundamentalEquivalence}

	{Consider a network of $N$ spiking neurons with membrane potential (and possibly additional variables) $V_t$. Let $V_0$ be the initial state of the network, and assume that:}
	\begin{enumerate}
		\item {the probability distribution $Y_0$ of the \review{$N$-dimensional random variable }  corresponding to the first spike times of the network starting from $V_0$ if no interaction occurs meanwhile is known (this random variable is called the \emph{initialization random variable})}
		\item {the probability distributions of the \review{one-dimensional} reset random variables $(Y^{(i)})_{i=1\ldots N}$ and}
		\item  {the \review{one-dimensional} interaction random variables $(\eta_{ij})_{(i,j)=1\ldots N}$ conditionally on the last value of the countdown process are known and only depend on the state of the countdown process at the time of the spike.}
	\end{enumerate}

 {Then the probability distribution of the spike times of the original network is equal to the law of the spike times of the countdown process (or equivalently of the related countdown sequence) with initial distribution $Y_0$, reset random variables $(Y^{(i)})_{i=1\ldots N}$ and interaction random variables $(\eta_{ij})_{(i,j)=1\ldots N}$. }
\end{theorem}

\begin{remark}
	{Note that the assumption that the interaction random variable only depends on the state of the countdown variable at the time of the spike can be weakened, and as we show below, it might depend on some additional variable, but needs to be independent of the underlying membrane potential. }
\end{remark}

\begin{proof}
	{This theorem actually stems from the very definition of the countdown process and of the reset and interaction random variables. Indeed, because of assumption 1, the law of the first spike elicited by each neuron of the network if no interaction occurs is by definition given by $Y_0$. Therefore, the law of the time $t_1$ of the first spike elicited in the network (resp. the index of the related neuron) is equal to the minimal value of $Y_0$ (resp. the coordinate index $i_1$ where this minimum is reached), hence equal to the first spike time (resp. neuron index) obtained in the countdown process defined. \\
	The next spike time of neuron $i_1$ if it receives no spike occurs meanwhile has by definition the law of $Y^{(i_1)}$, which is also the probability distribution of the countdown process after the spike emission. All other neurons in the network will fire at a new value given by $\eta_{ij}$ by definition of these variables, and these are also the updated values of the countdown process, i.e.  after taking into account the spike emission. \\
	Therefore, by induction, the theorem is proved: the spike times of the network are equal in law to the spikes emulated by the countdown process (or the countdown sequence). }
\end{proof}

{We note the following straightforward Markov property \review{that follows from the specific assumptions of theorem \ref{thm:FundamentalEquivalence}}}
\begin{proposition}\label{pro:MarkovProperty}
	{Under the assumptions of theorem \ref{thm:FundamentalEquivalence}, the countdown process (resp. the countdown sequence) constitutes a Markov process (resp. chain).}
\end{proposition}

{Note that we did not prove that the countdown process existed, nor that it was a Markov process. The results of theorem \ref{thm:FundamentalEquivalence} and of proposition \ref{pro:MarkovProperty} rely on the assumption that we are able to identify the initialization probability distributions $Y_0$, the reset random variables $Y^{(i)}$ and the interaction random variables $\eta_{ij}$. This is a more difficult task that we will deal with in sections \ref{section:inhib} and appendices \ref{append:FurtherModels} and \ref{append:SynDelays} in some specific networks where those random variables can be explicitly derived.}

{One of the difficulties in event-based modeling of networks is what we call the \emph{avalanche} phenomenon. This phenomenon consists in the simultaneous occurrence of infinitely many spikes. When connections are instantaneous and excitatory, with big enough synaptic weights, the spike emission of a neuron can instantaneously elicit spikes from the neurons directly connected to it, which in turn can elicit spikes in their neighborhood. Such a chain reaction can result in an overflow in the network, and even make the first neuron spike again, which closes the loop: at a given instant, the neuron fires infinitely many spikes. This phenomenon is an artifact of the assumed instantaneity of spike transmission and of the absence of refractory period in the model, this is why we will be introducing both features in our models \review{in appendix~\ref{append:SynDelays}. These features are particularly important when excitation is involved.}}

\medskip



\section{Application to some simple types of neuron models}
\label{section:inhib}
In this section, we aim at building a Markov chain governing the spike times, and to this end identify the law of the two random variables necessary in the definition of the countdown process: the initialization, the reset and the interaction random variables. 

\subsection{The reset and initialization random variables}\label{ssect:Reset}
The reset random variable of the countdown process corresponds to the time to the next spike of the neuron that just fired a spike, if no spike occurs meanwhile. {For all integrate-and-fire neurons, with linear or nonlinear dynamics, when neuron $i$ fires a spike at time $t^{*}$, the membrane potential is reset at a deterministic value $V_r^{(i)}$. The new countdown value at this time hence corresponds to the time it takes for the neuron to fire another spike if nothing occurs meanwhile, and therefore has the law} of the first hitting time after $t^{*}$ to the threshold function $\theta(t)$  of the membrane potential process, starting from the reset voltage  namely:

\[Y^{(i)} \sim \inf \Big\{ t > 0 ; \;\; V^{(i)}(t+t^*) > \theta(t+t^*) \Big \vert V^{(i)}(t^{*}) = V^{(i)}_{r}\Big\}\]

{This random variable is known in closed-form, or can be easily computed only for linear integrate-and-fire neurons. This is one of the reasons why we restrict our study to this class of neurons. Note however that, formally, this expression is valid whatever the neuron intrinsic dynamics and whatever the threshold chosen. The limitation only arises from the technical complexity of finding the probability distribution of first hitting times of stochastic processes, which remains a great endeavour in stochastic processes theory and still does not have any satisfactory systematic analytical or even numerical solution beyond linear processes. Note also that this random variable depends on $t^*$ only if the membrane potential process $V^{(i)}(t)$ or if the threshold $\theta$ are non-stationary. }

\begin{remark}
When the parameters of the model are not stationary, for instance when the input current or the threshold are deterministic non-constant functions of time, the reset random variable depends on the last spiking time of each neuron. Therefore in that case, in order to define the reset random variable at each spike time, one has to know the absolute time which can be seen as the Markov chain defined by the cumulative sum of the spike times. The Markov chain of the countdown process therefore involves the additional variable $T$ which is updated as $T^{n+1}=T^n+\inf_{i=1\cdots N}{X^{(i)}_n}$. 
\end{remark}

The initialization random variable transforms the initial condition on the membrane potential into the first spike time for each neuron if no spike occurs meanwhile. This random variable can be easily computed from the first hitting time of the membrane potential process to the threshold. Let for instance $V_0$ be the initial condition of the membrane potential at time $t_0$. The initialization random variable has the law of the first hitting time:
\[(Y_0)^{(i)} \sim \inf \Big\{ t > 0 ; \;\; V^{(i)}(t+t_0) > \theta(t+t_0) \Big \vert V^{(i)}(t_0) = V^{(i)}_0\Big\}\]

\medskip

Now that these random variables are defined and their role understood, we address the case of the interaction random variables.


\subsection{Integrate-and-fire models without synaptic integration}
We start by considering instantaneous current-based interactions. In these cases, the reception of a spike results in an instantaneous current that modifies the membrane potential by an amount equal to the synaptic connectivity weight, and the input current is modeled as a Brownian motion. We consider two different types of intrinsic dynamics: the so-called \emph{perfect} integrate-and fire neuron and the \emph{leaky} integrate-and-fire neuron. The Brownian current noise model consists in assuming that we are in the application domain of the classical diffusion approximation. This approximation is widely used in computational neuroscience, and details can be found for instance in \cite{holden:76,ricciardi:77,tuckwell:88}\footnote{{Note that formally most of the derivations done in the present manuscript might be performed with the use of the discontinuous (point processes) noise $S_i$. However, the expressions of the random variables involved in the dynamics of the countdown process would be harder to express in closed-form, and numerical evaluations of such random variables are generally not as efficient as the powerful tools of the stochastic theory.}}. 

\subsubsection{Perfect integrate-and-fire neuron}
We start by considering the so-called \emph{perfect integrate-and-fire neuron} with external input and Brownian noise. Between two spikes, the membrane potential of the neuron $i$, denoted by $V^{(i)}(t)$ satisfies the equation:
\begin{equation}\label{eqperfect}
	\tau_i {dV^{(i)}}(t)= I_e^{(i)}(t) dt + \sigma_i dW^{(i)}_t.
\end{equation}
In this equation, $\tau_i$ denotes the membrane potential time constant, $I_e^{(i)}(t)$ the deterministic input current. The independent external noise sources are modeled by the $N$ Brownian motions $(W^{(i)})_{1\leq i \leq N}$ and  $\sigma_i$ are the related standard deviations. 

The equation can be integrated in closed form, conditionally to the fact that we know $V^{(i)}(t_0)$ the membrane potential at time $t=t_{0}$:
\begin{equation}\label{eq:PIFsolution}
	V^{(i)}(t) = V^{(i)}(t_{0}))+ \frac 1 {\tau_i} \int_{t_{0}}^t I_e^{(i)}(s) \, ds + \frac{\sigma_i}{\tau_i} (W^{(i)}_t-W^{(i)}_{t_0})
\end{equation}
\review{The neuron fires when its membrane potential reaches the threshold $\theta(t)$, after what the membrane potential is reset to a constant value $V_r^{(i)}$ and a spike is emitted. At this time, $t^*$, the membrane potential of all its neighbors $j\in \V(i)$ is instantaneously increased by the negative synaptic weight: $V^{(j)}(t^*) = V^{(j)}(t^{*-}) + w_{ij}$.
\begin{equation*}
	V^{(i)}(t^-) = \theta(t) \Rightarrow \begin{cases}
		V^{(i)}(t^*) = V_r^{(i)}\\
		V^{(j)}(t^*) = V^{(j)}(t^{*-}) + w_{ij}
	\end{cases}
\end{equation*}}

{Assume that } neuron $i$ fires at time $t^*$. From the result of section \ref{ssect:Reset}, the reset random variable of the related countdown process has the law of the first hitting time of the Brownian motion starting from $V_r^{(i)}$ to the boundary $\frac 1 {\sigma_i}  (\theta(t+t^*) - \int_{t^*}^{t^*+t} I_e^{(i)}(s) \, ds ) $. This law can be computed using different mathematical or numerical methods (see e.g. \cite{holden:76,ricciardi:77,plesser:99} or the review \cite{touboul-faugeras:07b}). One of the most efficient methods to solve this problem in a general setting amounts to solving a Volterra integral equation. It is the method we use in our simulations when the input and/or the threshold are non-constant functions of time. Else, we will be using the well-known formulas of the first hitting time of the Brownian motion to linear boundaries (see e.g. \cite{karatzas-shreve:87}). 

The interaction random variable is deduced from the effect of a presynaptic spike incoming at a synapse. If neuron $j$ receives an inhibitory spike from neuron $i$ at time $t^*$, its membrane potential $V^{(j)}$ is shifted by an amount equal to $w_{ij}$ (see formula \eqref{eq:PIFsolution} with initial condition at time $t^*$ : $V^{(j)}(t^*) = V^{(j)}(t^{*-}) + w_{ij}$): 
Therefore, it will never cross the threshold before time $X^{(j)}(t^*)$, and at this time, its value is: 
\[V^{(j)}(X^{(j)}(t^*)) = \theta\big( X^{(j)}(t^*)\big) + w_{ij}. \]
Starting from this initial value the membrane potential integrates the input and the noise, and {therefore will not spike at time $t^*+X^{(j)}(t^*)$ as expected if neuron $j$ did not receive a spike}. The additional time after time $t^*+X^{(j)}(t^*)$ before the next spike time of neuron $j$ is therefore independent of the past of the process, and has the law of the random variable:
\begin{multline*}
\tau_{ij}=\inf \Big\{t>0; \theta(X^{(j)}(t^*)+t^*) + w_{ij} +\int_0^t I_e^{(i)}(s+X^{(j)}(t^*)+t^*) \, ds \\ 
+ \sigma_i W^{(i)}_t = \theta(t+t^*+X^{(j)}(t^*)) \Big\},
\end{multline*} 
i.e. the law of the first hitting time of the standard Brownian motion to the boundary
\begin{equation*}
t\mapsto \frac 1 {\sigma_i} \Big(\theta(X^{(j)}(t^*) +t^* +t ) - \theta(X^{(j)}(t^*) + t^*) - w_{ij} -\int_0^t I_e^{(i)}(s+t^*+X^{(j)}(t^*)) \, ds \Big).
\end{equation*}
The interaction random variable is therefore simply $\eta_{ij}(X^{(j)}(t^*)) = X^{(j)}(t^*) + \tau_{ij}$. The interaction between neurons therefore only amounts to the addition of an independent random variable on the countdown value. In the case of stationary inputs and threshold, this random variable has the law of the first hitting time of a drifted Brownian motion starting from $0$ to the constant barrier $\vert w_{ij} \vert$, whose density reads (see e.g. \cite{karatzas-shreve:87}):
\begin{equation}\label{eq:pdfBrownDrift}
p^{(i,j)}(t) = \frac{|w_{ij}|}{\sigma_i\sqrt{2\pi t^3}}e^{-\frac{(w_{ij}-I_e\,t )^2}{2t\sigma_i^2}} \mathbbm{1}_{\mathbbm{R}_+^*}(t)
\end{equation}

In the case of stationary inputs, we can convince ourselves that the countdown process is an \emph{autonomous Markov process}. Indeed, the sequence $(X_n)_{n\geq 0}$ (respectively and equivalently $(X_t)_{t\geq 0}$), see section \ref{section:theoretical}, is a Markov chain (respectively a Markov process) with transitions given by the law of the reset and interaction random variables, which only depend on the parameters of the model.

\begin{proposition}\label{pro:Autonomous}
	If the input current and/or the threshold depend upon time, then the chain $(X_n,t_n)_{n \geq 0}$ where $t_n$ is the time of the last spike fired in the network (respectively and equivalently $(X_t,t)_{t\geq 0}$) is an autonomous Markov chain (respectively	process).	
\end{proposition}

\begin{proof}
	\review{Indeed, we have seen that in that case both the reset and the interaction random variables depend on the 
	last spike time, and that the interaction random variable depends on $X^{(j)}(t^*)$. More precisely, let $(X_n,t_n)$ be the state of the Markov chain after $n$ spikes ($X_n$ is the $N$-dimensional countdown process and $t_n$ the sequence of spike times). Define $i_n = {\rm Argmin}_{i}\{X^{(i)}_n\}$. The neuron $i_n$ fires the $(n+1)^{\rm{th}}$ spike at time $t_{n+1} = t_n + X^{(i_n)}_n$. Its countdown value $X^{(i_n)}_{n+1}$ is reset by drawing from the law of the first hitting time of the standard Brownian motion starting from $V_r^{(i_n)}$ to the boundary 
	\[ t\mapsto \frac 1 {\sigma_{i_n}} \left ( \theta (t+t_n+X^{(i_n)}_n ) - \int_{0}^{t} I_e^{(i)} (t_n+X^{(i_n)}_n+s) \, ds \right),\] 
	and the state of the neurons $j$ connected to $i_n$ is updated by adding an independent random variable $\eta_{i_nj}$ having the law of the first hitting time of the standard Brownian motion to the boundary 
	\[t\mapsto \frac 1 {\sigma_j} \left (\theta(X^{(j)}_n + t_n + t ) - \theta(X^{(j)}_n + t_n ) - w_{ij} -\int_0^t I_e^{(i)}(s+X^{(j)}_n + t_n ) \, ds \right).\]}
	
	\review{Therefore, the transition probability only involve the value of the chain at rank $n$, since the formulas only depend on $t_n$ and $X_n$ and are independent of $(t_k, X_k)_{k<n-1}$ the previous values of the spike times or countdown values at previous ranks.} 
	
	We conclude from this observation and by application of theorem \ref{thm:FundamentalEquivalence}, that the sequence $(t_n,X_n)_n$ constitute a Markov chain the simulation of which provides access to the spike times of the network.
	
\end{proof}

\review{In all this section we considered time dependent thresholds $\theta$. This dependency upon time is dropped in the rest of the paper, since most of the classical integrate-and-fire neurons do not include this dependency. Note however that all the following developments can be performed using a time-dependent threshold in a same fashion as done here. }
\subsubsection{Leaky integrate-and-fire models}
\label{ssection:LIF}
\paragraph{The model}
We now take into account the leak of the membrane potential but no synaptic integration. The membrane potential is governed by equation:
\[
\begin{cases}
	\tau_i dV^{(i)}(t) &= (-V^{(i)}(t) + I_e^{(i)}(t))dt + \sigma_i dW_t^{(i)} \\
	V^{(i)}(t^-) &= \theta \Rightarrow V^{(i)}(t) = V_r^{(i)}
\end{cases}
\]
where the $(W^i_t)_{1\leq i \leq N}$ are independent Brownian motions. 
Therefore, the membrane potential $V^{(i)(t)}$ in the absence of correlation is an Ornstein-Uhlenbeck process. Between two spikes, the membrane potential can be written as:
\begin{equation}\label{eq:BaseLIF}
V^{(i)}(t) = V^{(i)}(t_{0}) e^{-(t-t_{0})/\tau_{i}} + \int_{t_{0}}^{t} e^{(s-t)/\tau_{i}} I_{e}^{(i)}(s) \,ds + 
\sigma_{i} \int_{t_{0}}^{t} e^{(s-t)/\tau_{i}} \, dW^{(i)}_{s}
\end{equation}
together with the spiking condition: 
\begin{equation*}
	V^{(i)}(t^-) = \theta \Rightarrow 
\begin{cases}
	V^{(i)}(t) &= V_r^{(i)}\\
	V^{(j)}(t) &= V^{(j)}(t^-) + w_{ij} \qquad \forall j \in \V(i)
\end{cases}
\end{equation*}
Let $t^*$ be the time when the neuron $j$ receives a spike from neuron $i$. We define $V^{(j)}_*:=V^{(j)}(t^{*-})$ and $X^{(j)}_*:=X^{(j)}(t^{*-})$. We have:
\begin{multline}\label{eq:diffLIF}
	V^{(j)}(t^* + t) = (V^{(j)}_* + w_{ij})e^{-t/{\tau_j}} + \frac 1 {\tau_j} \int_0^t e^{(s-t)/{\tau_j}}I_e^{(j)}(s+t^*)\,ds \\ 
	+\frac {\sigma_j} {\tau_j} \int_0^t e^{(s-t)/{\tau_j}}  \,dW_s^{(j)}
\end{multline}
Let $\widetilde{V}^{(j)}$ the membrane potential of the neuron $j$ if it did not receive the incoming spike at time $t^*$. It satisfies the equation:
\begin{equation*}
   \widetilde{V}^{(j)}(t^* + t) = V^{(j)}_* e^{-t/{\tau_j}} + \frac 1 {\tau_j} \int_0^t e^{(s-t)/{\tau_j}}I_e^{(j)}(s+t^*)\,ds +\frac {\sigma_j} {\tau_j} \int_0^t e^{(s-t)/{\tau_j}}  \,dW_s^{(j)}
\end{equation*}
Therefore we have:
\begin{equation*}
V^{(j)}(t^* + t) = \widetilde{V}^{(j)}(t^* + t) + w_{ij}e^{-t/{\tau_j}}
\end{equation*}
(note that this property simply stems from the linearity of the LIF neuron). For $t=X^{(j)}_*$ we have $\widetilde{V}^{(j)}(t^* + X^{(j)}_*) = \theta$ and, from \eqref{eq:diffLIF}, we have:
\begin{equation*}
V^{(j)}(t^* + X^{(j)}_* + t) = (\theta + w_{ij}e^{-X^{(j)}_*/{\tau_j}}) e^{-t/{\tau_j}} + \frac 1 {\tau_j} \int_0^t e^{(s-t)/{\tau_j}}I_e^{(j)}(s+t^*+X^{(j)}_*)\,ds + \frac {\sigma_j}{\tau_j} \int_0^t e^{(s-t)/{\tau_j}}  \,dW_s^{(j)}
\end{equation*}
We therefore deduce that the next spike time, i.e. the next hitting time of the barrier $\theta$ by the process $V^{(j)}$, conditionally to the random variable $X^{(j)}_*$ is the sum of $X^{(j)}_*$ and an independent random variable whose law is equal to the hitting time to the barrier $\theta$ of the process \eqref{eq:BaseLIF} with initial condition $V^{(j)}(0) = \theta + w_{ij} e^{-X^{(j)}_*/{\tau_j}}$ and with the time shifted input current $\tilde{I}_e^{(j)} (t) := I_e^{(j)}(t + t^* + X^{(j)}_*) $.
\[
\eta_{ij} \sim  \inf \left \{ t>0; \;\; U^{(j)}(t) = \theta \vert U^{(j)}(0) = \theta + w_{ij} e^{-X^{(j)}_*/{\tau_j}} \right \} + X^{(j)}_{*}
\]
where $U^{(j)}(t)$ is the solution of equation \eqref{eq:BaseLIF} with the  specified time-shifted current.

The problem of the first hitting time of the LIF neuron with constant or curved boundaries was addressed in many articles (see e.g. \cite{plesser:99,touboul-faugeras:07b} for a review and references therein), no closed-form solution is available, and this distribution can be efficiently computed by solving a Volterra equation, for instance. 

An important remark is that this random variable only depends on $X^{(j)}_*$. Conditionally to $X^{(j)}_*$, the added random variable is independent of the past of the process, so the sequence $X^{(j)}$ is Markovian. Furthermore, the network countdown process dynamics is autonomous: we do not need to refer to the underlying membrane potential to describe its evolution. This is very interesting since we can study and simulate this random variable by itself. Therefore, the variable $(X_t)$, possibly augmented with the time $t$ if the input current or the threshold are not stationary, is a Markov process, and this process sampled at the times of the spikes is a Markov chain. Furthermore, the law of the zeros of this process is equal to the one of the spikes of the underlying network. 

\paragraph{Event-based simulations}
The Markov chain we built provides a very natural event-based simulation algorithm for inhibitory stochastic networks. The method consists in simulating the Markov chain describing the time of the spikes for each neuron. We have seen that simulating the times of the spikes was equivalent in law to simulating the membrane potential, from the spikes viewpoint (see theorem \ref{thm:FundamentalEquivalence}). The event-based simulation consists in building this Markov chain. Simulating this Markov chain requires to draw at each spike time from the laws of  the reset and the interaction random variables. We have seen that these random variables can be expressed in most of the cases as first hitting times of random processes. In the cases where these laws are known in closed form, a very efficient simulation procedure can be used. If it is not the case, then we will have to evaluate these random variables. Assume that at the initial time $t_0$ the values of membrane potential of each neuron and of the additional variables of the model are known. The initial countdown value for a given neuron will be simply computed as the first hitting time of its membrane potential process to  the threshold, starting from this initial condition and can therefore be computed in the same way as the initialization variable. From this initial time, the principle of the algorithm is to build the discrete-time Markov chain containing as a variable the countdown process that gives the times of the spikes (we have seen that sometimes additional variables were necessary). Then to deduce the state of the chain at time $n+1$ knowing its state at time $n$, we use the recursion relation described in previous sections.
%
%

\begin{itemize}
 \item We first identify the neuron having the lowest countdown value, which amounts to finding the minimal value in a list of $N$ elements, an elementary operation efficiently coded. This neuron is the one that elicits the first spike.
 \item When this neuron is identified, we move the clock forward to this time, and draw the new state of the network: the neuron that just fired a spike is reset by drawing from the law of the related reset variable and the other neurons states are updated by drawing from the law of their respective interaction variables. Once the states of all neurons have been updated, the simulation proceeds.
\end{itemize} 

\begin{figure}
	\centering
		\subfigure[Spike time distributions, neuron 1]{\includegraphics[width=.30\textwidth]{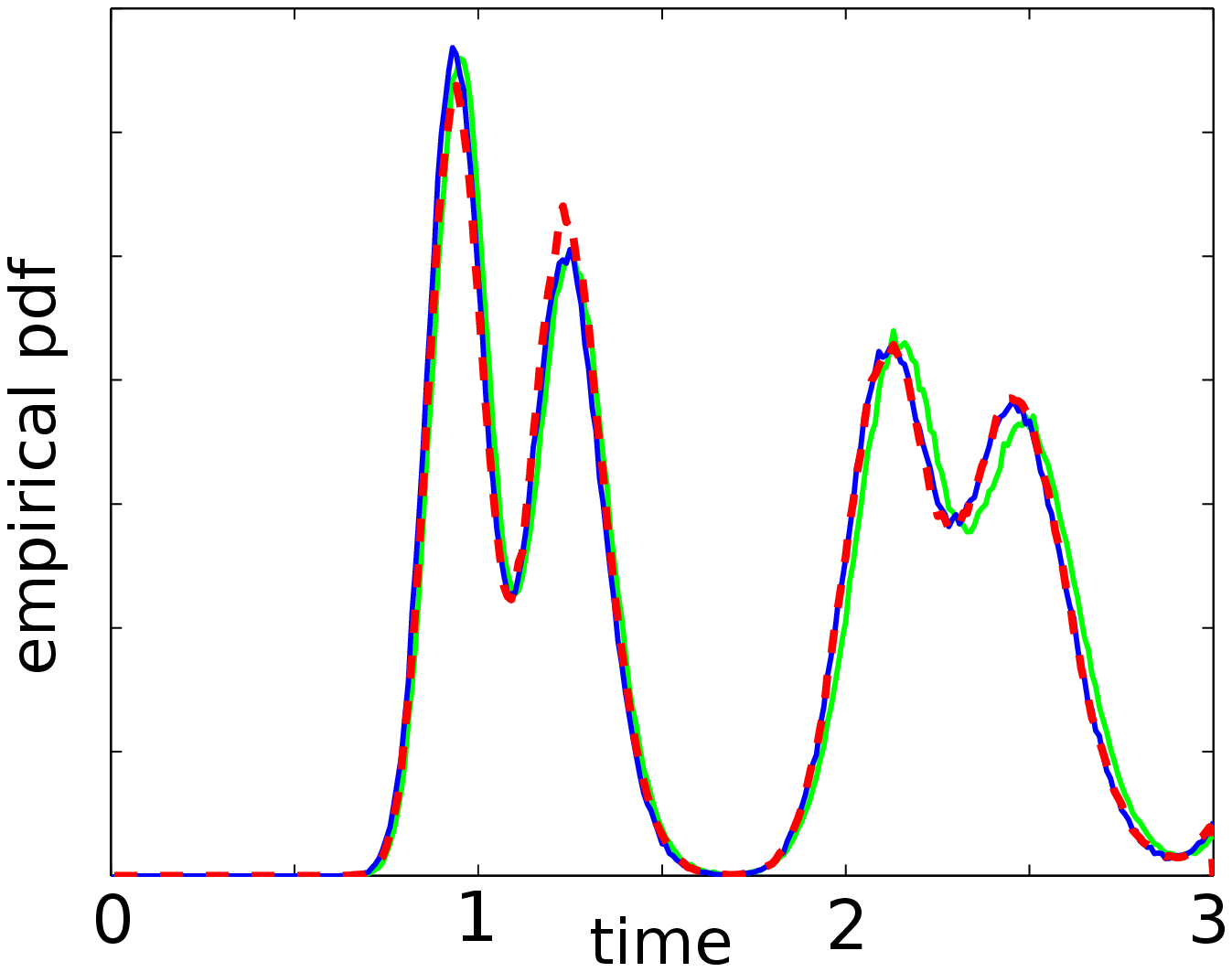}}\qquad
		\subfigure[Spike time distributions, neuron 2]{\includegraphics[width=.30\textwidth]{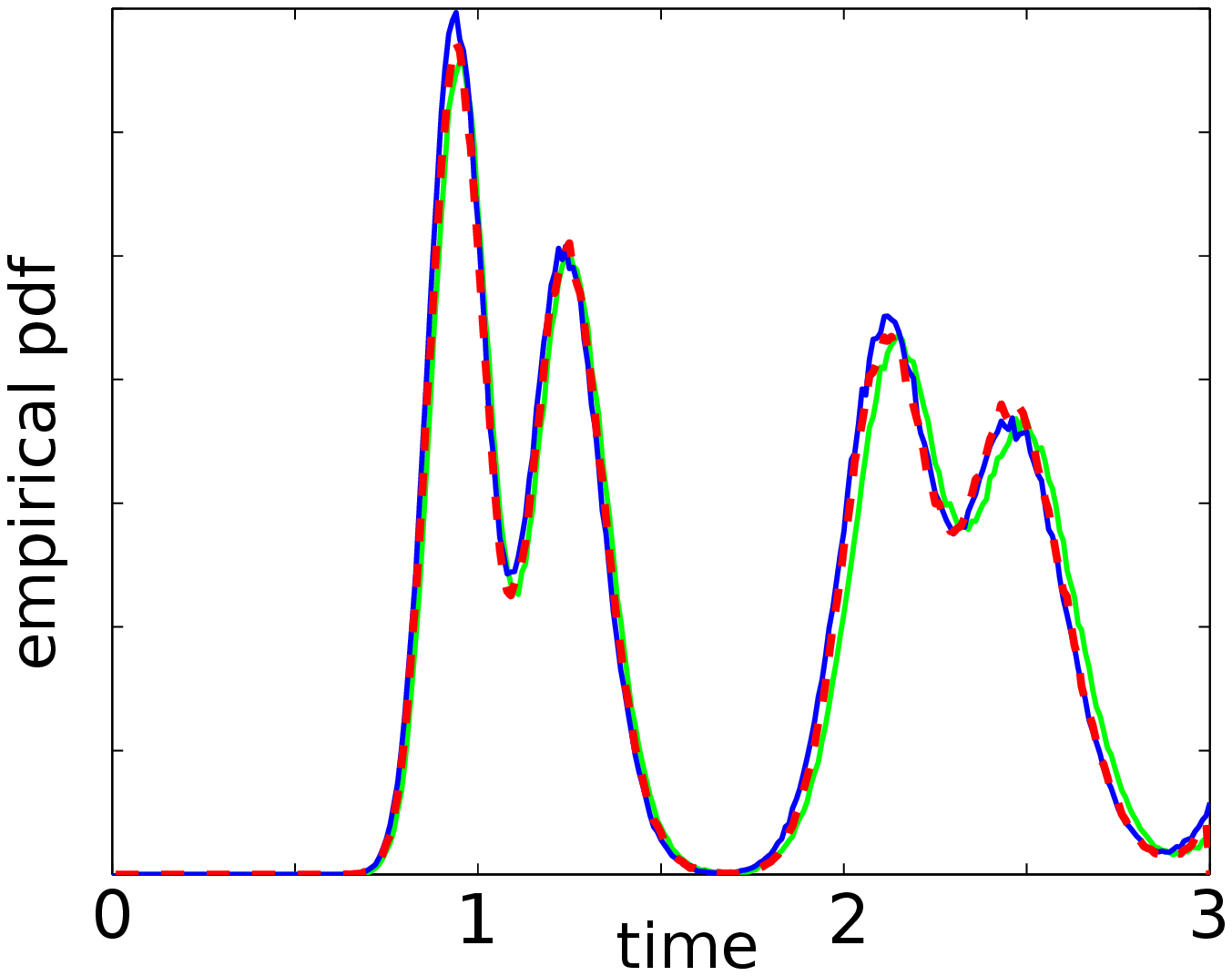}}\\
		\subfigure[Spike time distributions, neuron 1]{\includegraphics[width=.30\textwidth]{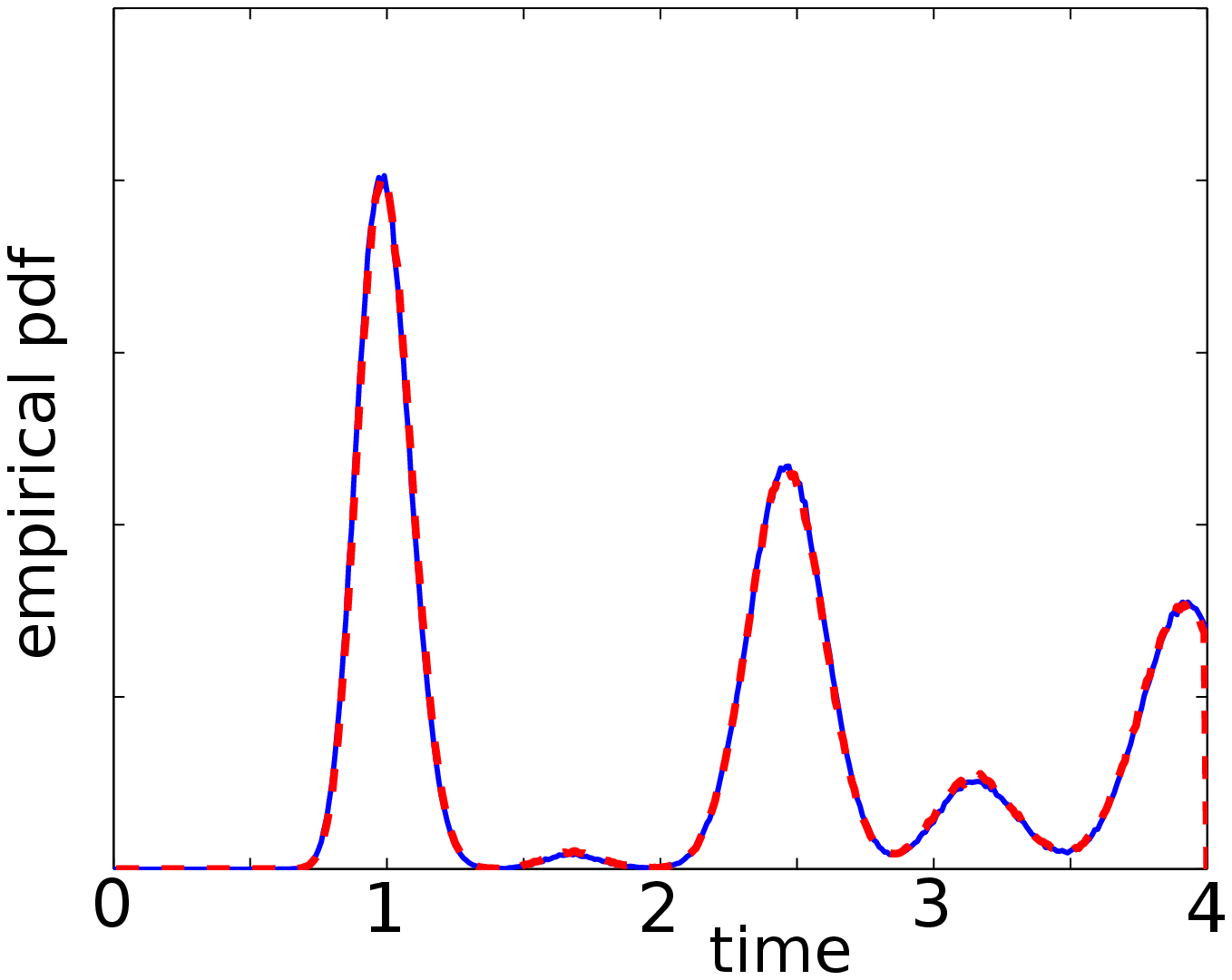}}\qquad
		\subfigure[Spike time distributions, neuron 2]{\includegraphics[width=.30\textwidth]{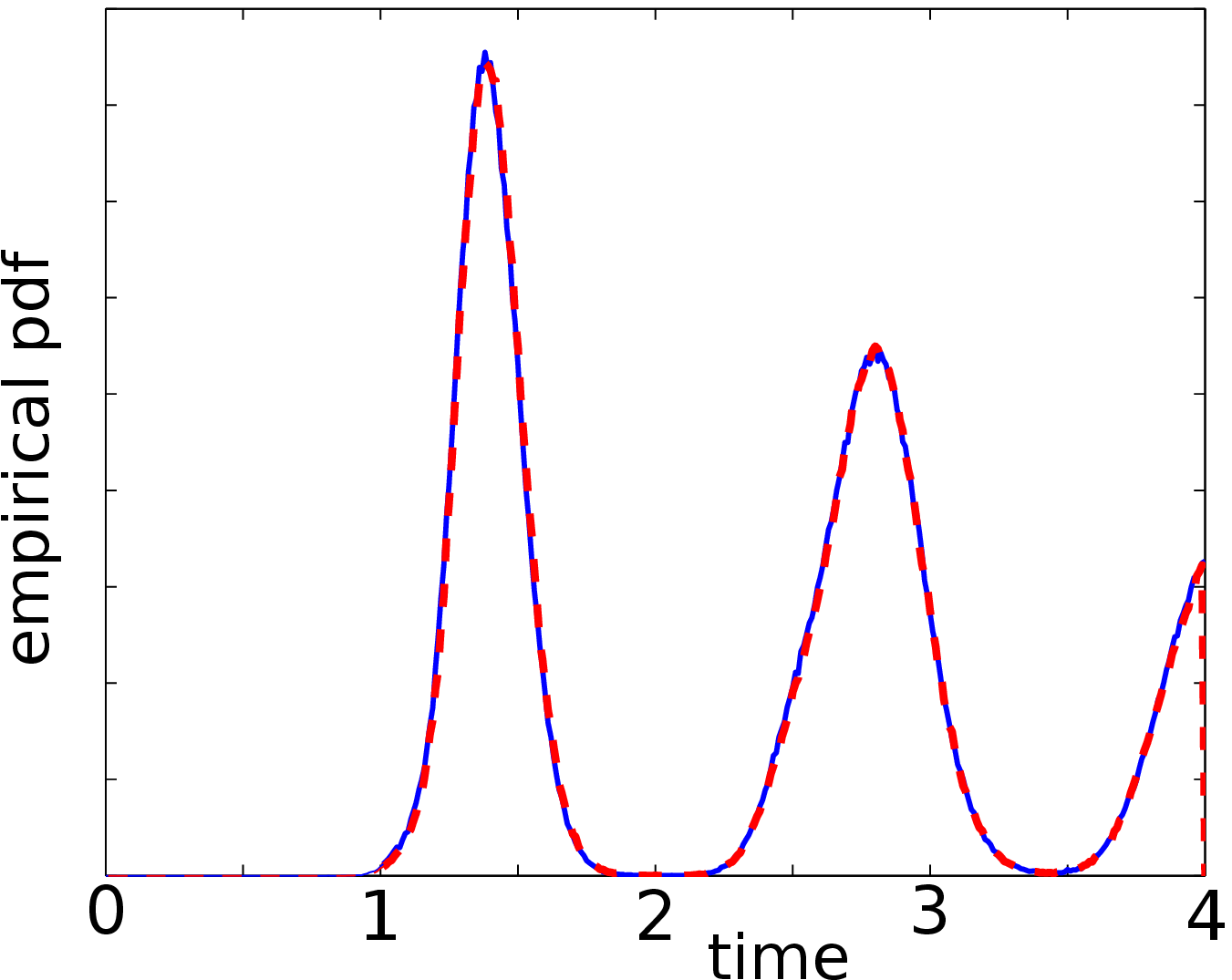}}\\
		\caption{\review{Spiking probability as a function of time for a fixed initial condition}: histogram computed over $500\,000$ realizations of the process. Comparison of the spike time distributions simulated by a Monte-Carlo method (green solid line), Gobet's ~\cite{gobet:00} Monte-Carlo improved method (blue solid line) and by our event-based simulation method (dashed red line). Left: Neuron 1, Right: Neuron 2, top row: symmetrical network with input current $I_1=I_2=1$, reset $V_r^1=V_r^2=0$, thresholds $\theta_1=\theta_2=1$ and connectivity weights $w_{12}=w_{21}=-0.2$. Bottom Row: Asymmetrical network, input current $I_1=I_2=1$, reset $V_r^1=V_r^2=0$, thresholds $\theta_1=1$, $\theta_2=1.3$ and connectivity weights $w_{12}=-0.1$ and $w_{21}=-0.5$. Note that because of the prohibitive execution time of Monte-Carlo methods, these are coded in C++ for the purpose of this simulation over a large number of sample paths. The time interval chosen is $[0,4]$ and the time step for Monte-Carlo algorithms is  $dt=0.01$. }
	\label{fig:Comparison}
\end{figure}

Let us for instance compare the natural Monte-Carlo simulation and the event-based simulation, in terms of precision and computational efficiency, in the case of a simple two-neurons perfect integrate-and-fire network model with instantaneous currents, each neuron having a constant threshold $\theta_i$ and receiving a constant input $I_i$. In that case, based on the previous analysis, the reset and the interaction random variables involved in the dynamics of the countdown chain respectively have the following densities:
\begin{equation}\label{eq:densityPIF}
	\begin{cases}
		p_Y^i(t) = \frac{\theta_i-V_r^i}{\sigma_i\sqrt{2\,\pi\,t^3}} \exp\Big(-\frac{(\theta_i-V_r^i-I_i\,t)^2}{2\,t\sigma_i^2}\Big)\\
		p^{ij}(t) = \frac{\vert w_{ij}\vert }{\sigma_i\sqrt{2\,\pi\,t^3}} \exp\Big(-\frac{(w_{ij}+I_i\,t)^2}{2\,t\sigma_i^2}\Big)\\
	\end{cases}
\end{equation}
In figure \ref{fig:Comparison} we compare the results of the simulations of the two algorithms on $500\,000$ realizations of the network and observe as predicted a very good match between the two distributions. In order to directly simulate the sample paths and the spike times, Monte-Carlo methods prove relatively inefficient, since the error in the spike time is of the order $\sqrt{dt}$ if $dt$ is the time step of construction of the sample path (see \cite{gobet:00}). In order to improve the accuracy of the algorithm, Gobet in \cite{gobet:00} 
proposed the interesting idea to interpolate the threshold crossing using the law of the maximum of a Brownian bridge conditioned on the values it takes at the discretized times. For the event-based algorithm, we tabulated the functions given in equations \eqref{eq:densityPIF} for generality\footnote{The case of \review{PIF neurons} is the only case where both the interaction and the reset random variable can be expressed in closed form with usual functions.}. We use both natural and Gobet's refined Monte-Carlo method and compare the precision and the execution time of the numerical algorithms. Implemented in Matlab\textregistered R2008b, and run on a MacBook 2Ghz Interl Core 2 Duo with 2GB 1067 MHz DDR3 memory, Gobet's enhanced Monte-Carlo algorithm for $50\,000$ sample paths runs in 716.68 seconds. On the same machine, we simulated the natural Monte-Carlo algorithm using the same discretization time-step, in which case the precision is not at the level of both Gobet's and the event based algorithm as we observe in Figure \ref{fig:Comparison}, the execution time is of 19.04 seconds. Note that in order to achieve the same precision on the spike times as Gobet's method with discretization time step $dt$, one should use a time step of the order of $dt^2$ for the regular Monte-Carlo method, in which case the execution time of the event-based algorithm for the same parameters shall be a hundredfold larger, a prohibitive execution time for such a simple model in such small dimensions. On the same machine and with the same number of sample paths, the event-based algorithm runs in 2.46 seconds. Simulating the event-based algorithm in this case therefore yields a gain of the order of 30. It therefore clearly appears that the event-based algorithm is faster. Moreover, increasing the precision of the latter only involves a more precise tabulation and does not affect the efficiency of the algorithm, whereas dividing for instance by 10 the time step of a Monte-Carlo algorithm multiplies by 10 the execution time. Note also that the number of neurons in the network and the time interval over which the simulation is performed will not modify this ratio. 

\review{Let us now describe the result of figure Fig.~\ref{fig:Comparison}. In the top row we show the case of a symmetrical network. In the absence of noise and interaction, each neuron fires with period 1. The presence of noise results in the fact that the distribution of the first spike time will have its median value at $1$ with a certain spread around this value. Therefore one of the neurons (say neuron 1 without loss of generality) will fire first. It instantaneously inhibits neuron 2 that will now fire after some delay. Because of the symmetry of the network, this phenomenon accounts for the multimodal nature of the distribution. Consecutive spikes reproduce the same scheme. The asymmetrical case can be understood in the same fashion: neuron 1 fires with a median ISI equal to 1, and neuron 2 with median ISI 1.3. Neuron 1 fires more frequently, but its connectivity weight is smaller and affects less the state of neuron 2 than neuron 2 affects the state of neuron 1. These phenomena result in shifting the peaks of emitted spike times accordingly. } 

\medskip

Therefore, we have verified that in the case of instantaneous synapses, the sequence of spike times constitutes a Markov chain, whose transition probability was explicitly derived from the law of the first hitting time of the membrane potential process to some boundary. This result extends to cases where on takes into account synaptic integration as we now show. In appendixes  \ref{append:FurtherModels} and \ref{append:SynDelays}, we show that these results can be extended to more complex cases, but fail to do so for conductance-based models.

\subsection{Integrate-and-fire models with exponentially decaying synaptic currents}
In this section we consider that the spike integration and the noisy integration are not instantaneous. {Note that by exponentially decaying synaptic currents, we are only concerned with the noise-generated current, and we will consider in this section both instantaneous and exponentially decaying spike integration}.   We start by considering, similarly to the previous case, the perfect integrator case first before addressing the case of the leaky integrator.
\subsubsection{Perfect integrate-and-fire model} 
A different way for improving the perfect integrate-and-fire model is to consider that the inputs received by the neuron $i$ are integrated at the level of the synapse with a characteristic time constant $\tau_s^{(i)} \neq 0$. In that case, the membrane potential satisfies the equations:
\[
 \begin{cases}
  \der{V_t^{(i)}}{t} &= I_e^{(i)}(t) + I_s^{(i)}(t) \\
  \tau_s^{(i)} dI_s^{(i)}(t) &= -I_s^{(i)}(t) dt + \sigma_i dW^{(i)}_t
 \end{cases}
\]
whose solution reads:
\begin{multline}\label{eq:PIFSI}
 V^{(i)}(t) = V^{(i)}(t_0) + \int_{t_0}^t I_e^{(i)}(u)\,du +\tau_s^{(i)}(1-e^{-t/\tau_s^{(i)}}) I_s^{(i)}(t_0)  +\\
  \sigma_i \int_{t_0}^t \int_{t_0}^v e^{-(v-u)/\tau_s^{(i)}}\,dW_u^{(i)}\,dv,
\end{multline}
or in the leaky integrate-and-fire case
\begin{equation}\label{eq:LIFEDSP}
\begin{cases}
 \tau_i dV^{(i)}(t) &= (\mu_i - V^{(i)}(t)) dt +I_e^{(i)}(t) dt + I_s^{(i)}(t) dt \;\;\;\; 1\leq i \leq N \\
 \tau_s^{(i)} dI_s^{(i)}(t) &= -I_s^{(i)}(t) dt + \sigma_i dW^i_t 
\end{cases}
\end{equation}
in which case the membrane potential reads
\begin{equation*}
 V_t^{(i)} = V^{(i)}(0) + \int_{0}^t I_e^{(i)}(s)\,ds +\tau_s^{(i)}(1-e^{-t/\tau_s^{(i)}}) I_s^{(i)}(0) + \sigma_i \int_0^t \int_0^s e^{-(s-u)/\tau_s^{(i)}}\,dW_u^{(i)}\,ds,
\end{equation*}
and a spike is emitted when the membrane potential reaches a threshold $\theta$ which we assume constant. 

We now consider the related countdown process $(X_t)_t$ and compute the random variables that are necessary to the definition of its dynamics. The related reset random variable in the case of the exponentially decaying synaptic currents has the law of the first hitting time of the Doubly Integrated Process (DIP) $\int_0^t \int_0^v e^{-(v-u)/\tau_s^{(i)}}\,dW_u^{(i)}\,dv$ or Integrated Wiener Process (IWP) $\int_0^t W_u^{(i)} \,du$ to a curved boundary depending on the inputs of the neuron and the initial condition of the synaptic input. These hitting times can be evaluated based on methods proposed e.g. in \cite{mckean:63,goldman:71,lachal:91,lachal:96b} in the case of constant input and constant threshold, or approximated using the framework developed in \cite{touboul-faugeras:08}. Note that in this latter case no closed-form solution can be provided except in the case of the IWP  with a piecewise quadratic polynomial input.

Let us now compute the interaction random variable, in the cases of instantaneous or exponentially decaying synaptic spike-integration currents.

\paragraph{Instantaneous synapses}
In the case of instantaneous synaptic integration of the spikes, the linearity of the equation yields the fact that the interaction random variable has the law of the first hitting time of the threshold $\theta$ of the membrane potential process starting from $(\theta+w_{ij}e^{-X^{(j)}_*/\tau_{s}^{(j)} }, I_s^{(i)}(X^{(j)}_*)))$ to reach the threshold $\theta$, and can therefore be computed using the same approximations or formulas depending on the model we choose and the type of input current considered. Furthermore, because of the strong Markov property of the membrane potential process, the interaction random variable is independent of the trajectory of the voltage and only depends on $X^{(j)}_*$.

\paragraph{Synaptic integration}
Taking into account synaptic integration in the connections makes more sense: it corresponds to the assumption that whatever its origin, the inputs received at the level of the synapse are integrated in the same fashion. In that case the effect of an incoming spike on a postsynaptic neuron is added instantaneously to the synaptic current. Therefore, using the same technique as before, we can obtain the law of the interaction variable. For the perfect integrate and fire neuron with exponentially decaying synaptic current, if neuron $j$ receives an incoming spike from neuron $i$ at time $t^*$, the law of this random variable is deduced from the law of the first hitting time of the related DIP starting from $\left(\theta+w_{ij}\tau_s^{(j)}(1-e^{-X^{(j)}_*)/\tau_s^{(j)}}), I_s\left(t^{*}+X^{(j)}_*\right)+w_{ij}e^{-X^{(j)}_*/\tau_s^{(i)}}\right)$ to reach the threshold $\theta$. 
In the case of the perfect integrate-and-fire neuron with perfect synapses, the law of the interaction random variable is given by the law of the first hitting time of the related IWP starting from $(\theta+X^{(j)}_* w_{ij}, I_s(X^{(j)}_*)+w_{ij})$ to reach the threshold $\theta$, and this random variable is independent of the membrane potential and only depends on $X^{(j)}_*$.

\paragraph{Building a Markov chain}
In these two cases the countdown process is not a Markov process. If we  consider in addition the value of the synaptic current $I_s(t_{n})$ at the time $t_{n}$, we obtain a Markov chain modeling the times of the spikes. More precisely, consider $(X_n, I_s(t_{n}), t_n)$ the countdown process together with the synaptic current at the next spike time and $t_n$ the last spike time. The first spike will be fired by the neuron $i_n$ having the lowest countdown value. It will fire at time $t_{n+1}=t_{n}+X^{(i_n)}_n$. Its countdown value and the value of the synaptic current at the next spike time is computed by drawing from the law of an independent pair composed of the next spike time and the relative location of the synaptic currents at this time (see \cite{touboul-faugeras:08}). The countdown value and the future synaptic current of each neuron $j\in \V(i)$ is updated by drawing from the law of the first hitting time of the membrane potential of neuron $j$, and the other neurons states are unchanged. This chain is a Markov chain whose spike times are equal in law to spike times of the original network, by direct application of theorem \ref{thm:FundamentalEquivalence} and of proposition \ref{pro:MarkovProperty}.

\subsubsection{Leaky integrate-and-fire model}
Our last example features both a leaky intrinsic dynamics and synaptic integration of the noise. In this case the membrane potential and the synaptic noise are coupled via the differential equation \eqref{eq:LIFEDSP} with the spiking condition given by:
%
\begin{equation}\label{eq:EDSP}
V^{(i)}(t^-) = \theta \Rightarrow 
\left \{ 
\begin{array}{lll}
V^{(i)}(t) &=& V_r^{(i)}\\
I_s^{(j)}(t) &=& I_s^{(j)}(t^-) + w_{ij} \mathbbm{1}_{j \in \V(i)}
\end{array}
\right .
\end{equation}
Qualitatively, when a spike is received by a neuron, the synaptic current $I_s$ integrates the spike and the effect on the membrane potential is smoother. Therefore in this model it is interesting to consider post-synaptic pulses having the same dynamics as the noise integration, i.e. solution of the differential equation:
\[\tau_s^{(i)} \der{I_a^{(i)}}{t} = -I_a^{(i)}(t).\]
The very same analysis could be done if we considered an instantaneous spike integration. We omit it here since it follows in a \review{straightforward} fashion from the upcoming analysis of a slightly more complicated case.

The reset random variable is given by the first hitting time of the membrane potential stochastic process. 
The same type of calculations as in the previous section yields the relation: 
\begin{equation}\label{eq:EDSPSpikeDiff}
V^{(j)}(t^* + t) = \widetilde{V}^{(j)}(t^*+t) + e^{-t/\tau_j} w_{ij} \frac{1-e^{-\alpha_j t}}{\alpha_j},
\end{equation}

\noindent for $j \in \V(i)$ and $\tau_j \neq \tau_s^{(j)}$. $\alpha_j = \frac 1 {\tau_s^{(j)}} - \frac 1 {\tau_j}$ and again $\widetilde{V}^{(j)}(t^*+t)$ is the membrane potential of the neuron $j$ without any interaction.  We can see that after the time $X^{(j)}_*$, the membrane potential of $j$ is $\theta + w_{ij}e^{-t/\tau_j}\frac{1-e^{-\alpha_j X^{(j)}_*}}{\alpha_j}$. The evolution of the potential $V^{(j)}$ after $t^* + X^{(j)}_*$ and conditionally on $X^{(j)}_*$ and $I_s(t^*)$ is independent of the past, so we have to wait for the process \eqref{eq:EDSP} to reach the threshold $\theta$ from the initial condition $\theta + w_{ij}e^{-t/\tau_j}\frac{1-e^{-\alpha_j X^{(j)}_*}}{\alpha_j}$ and with the ``time and space'' shifted currents $\tilde{I_e^j}(t) := I_e^j(t + t^* + X^{(j)}_*) + w_{ij}e^{-\frac{X^{(j)}_*}{\tau_s^{(j)}}}$. 

In the case $\tau_j = \tau_s^{(j)}$ we replace the expression $\frac{1-e^{-\alpha_j X^{(j)}_*}}{\alpha_j}$ by $X^{(j)}_* w_{ij} e^{-t/\tau_j}$, and the change in the currents is the same.

Therefore, the variable $(X_t, I_s(t))$ is Markovian and we deduce the precise firing times from its study. This Markovian variable requires to evaluate the law of the first hitting time of a DIP to a curved boundary, which can be achieved using the technique developed in \cite{touboul-faugeras:08}. As in the case of the perfect integrate-and-fire neuron, we can show that this process, possibly augmented with the time process $t$, and possibly sampled at the times of the spikes, satisfies the Markov property and the law of the zeros of the Markov process is the same as the law of the spikes of the underlying network. 

In the case of non-stationary inputs, the dynamics of the countdown process together with the synaptic current at the next firing time and the last spike time can be described as follows: let us note $(X_n, I_{s,n}, t_n)$ the previous Markov process  after the $n^{\rm{th}}$ spike. The next spike will be fired by the neuron $i_n$ having the lowest countdown value. It will fire at time $t_{n+1}=t_{n}+X^{(i_n)}_n$. Its countdown value will be reset to the first hitting time of the related DIP, which can be evaluated as before. Therefore, by drawing from the law of this pair, we have  the new countdown value and the future synaptic current at the time of the next spike for $i_n$. Similarly, each neuron $j\in \V(i_n)$ is updated according to the law of the first hitting time of the related membrane potential starting with an input current given by $I_{s,n}^{(j)}$ to reach a given threshold, and therefore the new countdown value and the future location of the input current are computed at the same time using the same results as before. The other neurons states are unchanged. It is clear that the law of the spikes is the same as the law of the zeros of the countdown process (see theorem \ref{thm:FundamentalEquivalence}).

\section{Excitatory Networks}\label{sec:Excitation}
We have shown in the previous section that when the network is inhibitory, the strong Markov property of the membrane potential process and the linearity of the equation allowed to define an event-based description of the network activity. In appendices \ref{append:FurtherModels} and \ref{append:SynDelays}, we show that these results can be extended to other more complex cases, but not to conductance-based models. In this section, we are interested in extending these results to the case of excitatory interactions. 

\reviewJ{As already mentioned, a well-posedness problem arises when taking into account excitatory interactions. This is due to the instantaneity of the interactions which  might lead to avalanche phenomena, i.e. self-excitation of the cell through the network at a given instant. A suitable way to avoid such problems is to take into account two important biological phenomena: the synaptic transmission delays and the refractory period. Indeed, both phenomena prevent the occurence of avalanches by upperbounding the firing frequency and forbidding instantaneous self-excitation. We show in appendix~\ref{append:SynDelays} that  they  can be introduced in the model and only amount to transforming the interaction random variables involved in the simple case where no delay or refractory period is taken into account. This is why, in the case of excitatory connections, we focus on the arrival of a single incoming excitatory spike at a given time, keeping in mind that, thanks to a suitable implementation, these two phenomena can be taken into account. } 

\subsection{\reviewJ{General facts}}
\reviewJ{We start by formally presenting the method on an unspecified general linear model of the same type as the ones covered in section~\ref{section:inhib} and some of the models of appendix~\ref{append:FurtherModels} where the Markovian description of the countdown process is possible. Let us assume that at time $t=t^*$, neuron $j$ receives an excitation coming from neuron $i$. Note that, strictly speaking, this scenario covers cases including synaptic delays and refractory period\footnote{Indeed, if the action of the spike fired by neuron $i$ at time $t_i$ affects neuron $j$ after a delay $\Delta_{ij}$, then $t^*=t_i+\Delta_{ij}$ and if $t_j$ is the last firing time of neuron $j$ and the refractory period corresponds to modulating the interaction by a function $\kappa$, then the synaptic weight $w_{ij}$ is multiplied by $\kappa(t^*-t_j)$. The way to include such phenomena in a generic manner is presented in details in appendix~\ref{append:SynDelays}.}.}

\reviewJ{The general principle of our approach for defining the future spike time of neuron $j$ is based on characterizing the value of the membrane potential (and of the additional variables needed to define the state of the membrane potential involved in its dynamics) of neuron $j$ at time $t^*$. Let us first treat the unidimensional cases of the PIF and LIF neurons with instantaneous synaptic currents. Let us assume for that both the last spike time $t_{\textrm{last}}$ and the value of the membrane potential  $V^{(j)}(t_{\textrm{last}})$ at this time are known. Then the law of the countdown value after spike reception can be written, using the total probability formula:  
\begin{multline}\label{eq:TotalProba}
	\cProb{X^{(j)}(t^{*+})\in dt}{X^{(j)}(t^*), t_{\textrm{last}}, V^{(j)}(t_{\textrm{last}})} \\= \int_{v_j\in\R}\cProb{{}^{j}\tau_{t^*,v_j+{w}_{ij}}^{\theta}\in t^* + dt}{V^{(j)}(t^*)=v_j} \cProb{V^{(j)}(t^*)\in dv_j}{ X^{(j)}(t^*), t_{\textrm{last}}, V^{(j)}(t_{\textrm{last}})}
\end{multline}
where ${}^{j}\tau_{s,u}^{\theta}$ is the first hitting time of the threshold $\theta$ of $V^{(j)}$ starting from initial condition $u$ at time $s$. We used the traditional notation in the stochastic literature $\mathbb{P}(\zeta \in dz)$, which represents the probability measure of a random variable $\zeta$ at $z$. This can be seen as the infinitesimal probability of $\zeta$ to be in $dz$. In all the cases we consider the random variables have densities with respect to the Lebesgue measure, and this quantity is equal to $p_{\zeta}(z)\,dz$ where $p_{\zeta}$ is the probability density of $\zeta$ with respect to the Lebesgue's measure $dz$. }

\reviewJ{Formula \eqref{eq:TotalProba} readily extends to multidimensional cases such as the bidimensional PIF and LIF neurons with synaptic integration treated in this manuscript. Indeed, assuming again that the last firing time $t_{\textrm{last}}$ and the values of the membrane potential and of the synaptic current at this time $(V^{(j)}(t_{\textrm{last}}), I_s^{(j)}(t_{\textrm{last}}))$ are known, the total probability formula yields for the probability of the next firing time:
\begin{multline}\label{eq:TotalProbaTwoD}
	\cProb{X^{(j)}(t^{*+})\in dt}{X^{(j)}(t^*), t_{\textrm{last}}, (V^{(j)}(t_{\textrm{last}}),  I_s^{(j)}(t_{\textrm{last}}))} \\= \int_{(v_j,i_j)\in\R^2}\cProb{{}^{j}\tau_{t^*,v_j,i_j+{w}_{ij}}^{\theta}\in t^* + dt}{V^{(j)}(t^*)=v_j,I_s^{(j)}(t^*)=i_j}\\ \times \cProb{V^{(j)}(t^*) \in dv_j, I_s^{(j)}(t^*) \in di_j }{ X^{(j)}(t^*), t_{\textrm{last}}, V^{(j)}(t_{\textrm{last}}), I^{(j)}_s(t_{\textrm{last}})}
\end{multline}
 }

\reviewJ{The first term in the integral of equations \eqref{eq:TotalProba} and \eqref{eq:TotalProbaTwoD} involves the probability distribution of the first passage time of the membrane potential conditionally to a fixed initial value, which corresponds exactly to the problem of finding the initialization random variable. As shown in section \ref{ssect:Reset}, this variable is directly related to first hitting times of the membrane potential process, and different ways for computing this law are discussed depending on the model considered throughout section~\ref{section:inhib}.}
\reviewo{The second term involves the probability distribution of the membrane potential (and possibly additional variables such as the synaptic current) conditioned on an initial state and on the value of its first hitting time to the threshold boundary. This law can be computed through the use of Bayes' formula for conditional probability.} \reviewNew{Let us treat for instance the case of a one-dimensional model (the general case is treated in a totally identical manner, but involves more intricate formulae). We denote for the sake of compactness of notations $\tau$ the first hitting time ${}^{j}\tau_{t^*,V^{(j)}(t^*)}^{\theta}$ of the process $(V^{(j)}_t)_{t\geq t^*}$ to the threshold $\theta$, with initial condition at time $t^*$ equal to $V^{(j)}(t^*)$. We have, using Bayes' formula for probability densities and the Markov property on $V^{(j)}$:
\begin{align}
	\nonumber \displaystyle{\cProb{V^{(j)}_{t^*}\in du}{ V^{(j)}_{t_\textrm{last}}, \tau=t^*+X^{(j)}_{t^*}}} &= \displaystyle{\frac{\cProb{\tau=t^*+X^{(j)}_{t^*}}{ V^{(j)}_{t^*}=u, V^{(j)}_{t_{\textrm{last}}}} \cProb{V^{(j)}_{t^*} \in du}{V^{(j)}_{t_{\textrm{last}}}}} {\cProb{\tau=t^*+X^{(j)}_{t^*}}{V^{(j)}_{t_{\textrm{last}}}}}}\\
	\label{eq:Bayes}		&= \displaystyle{\frac{\cProb{\tau=t^*+X^{(j)}_{t^*}}{V^{(j)}_{t^*}=u} \cProb{V^{(j)}_{t^*}\in du}{V^{(j)}_{t_{\textrm{last}}}}}{\cProb{\tau=t^*+X^{(j)}_{t^*}} {V^{(j)}_{t_{\textrm{last}}}}}}
\end{align}
}


\reviewNew{The formula \eqref{eq:TotalProba}, thanks to the relationship \eqref{eq:Bayes} only involves quantities that are already computed or easily derived. Indeed, it involves i) the law of the first hitting time of the voltage process to the boundary conditioned on its initial datum, problem which is exactly solved as the computation of the reset random variable in section~\ref{ssect:Reset}, and ii) the probability density of the Markovian transition of the voltage potential between $t_{\textrm{last}}$ and $t^*$. This transition is very easy to estimate from the properties of the underlying voltage process. Indeed, in all the cases treated, the voltage potential is a Gauss-Markov process, hence the transition probability is a Gaussian process whose mean and covariance matrix are very simple to compute as a function of the parameters of the membrane potential process. This implies that the characterization of the law of the first hitting time of the membrane potential to the spiking threshold allows characterization of both the inhibitory and the excitatory interaction random variable.}

\reviewNew{Formulae \eqref{eq:TotalProba} and \eqref{eq:TotalProbaTwoD} are compact formulations for the law of the next spike time of neuron $j$. It has an intricate form involving an integral that might be difficult to compute. However, we can express the interaction random variable for excitatory interactions in a more intuitive and efficient way using to use a two-step algorithm, providing a straightforward simulation method. Indeed, in order to compute the nest spike time, one only needs to draw a value of the next firing time from the law given by \eqref{eq:TotalProba} or \eqref{eq:TotalProbaTwoD}. And to complete this step, the computation of the integral involved in the definition is not necessary, since drawing a realization of this random variable can be more easily performed as follows:
\begin{itemize}
	\item first, draw the value $v_j$ (or $(v_j,i_j)$) from the law of the membrane potential process (and possibly additional variables) conditioned on initial datum and on the value of its first hitting time, computed through Bayes' formula on conditional probabilities \eqref{eq:Bayes}, or its multidimensional versions depending on the neuron model,
	\item second, draw the value of the first hitting time of the diffusion process $V^j$ to the boundary starting from initial condition $v_j+w_{ij}$ (or $(v_j,i_j+w_{ij})$).
\end{itemize}}
\reviewNew{If the probability distribution of the first hitting time of the membrane potential to the spiking threshold is well characterized, both random variable involved is also well defined. The characterization of such first passage time problems was treated in section~\ref{section:inhib}, and therefore allow event-based description of the spike times. We now make explicit the transition density in the cases treated in section~\ref{section:inhib}, and emphasize on the case of the PIF neuron since its simplicity allows compact and clear presentation. Other cases are less detailed since the analysis of the system and the Markov property are derived in a very similar fashion. }



\subsection{\review{PIF Neuron with Instantaneous Interactions}}
\reviewJ{Let us start by dealing with the case of the PIF neuron with instantaneous interactions. The dynamics of the membrane potential is given by equation \eqref{eqperfect}, in which case each membrane potential is an independent Brownian motion. Let $t_0$ denote the last spike time fired in the network and by $V^{(j)}_0$ the value of the membrane potential of $j$ at this time. Neuron $j$ is expected to reach the threshold $\theta$ and fire at time $t_1=t_0+X^{(j)}(t_0)$. Meanwhile, at time $t^*\in [t_0,t_1]$, the neuron receives an incoming spike from neuron $i$ that instantaneously shifts the membrane potential value by an amount equal $w_{ij}>0$. \reviewNew{Bayes's formula \eqref{eq:Bayes} allows characterizing the value of} $V^{(j)}(t^*)$ of the membrane potential of neuron $j$ at the time it receives the excitatory spike from neuron $i$. We have the following:} 

\begin{proposition}\label{prop:ExcitationPIF}
	\review{A network of perfect integrate-and-fire neurons with excitatory and inhibitory connections has a Markovian event-based description.}
\end{proposition}

\begin{proof}
\reviewJ{Let us denote \reviewo{as usual} by $(t_n)_{n\in \N}$ the sequence of spike times fired in the network and by $V_n=(V^{(k)}_{t_n}, k=1\ldots N)$ the values of the membrane potential of all the neurons at each time. We consider the sequence $Z_n=(t_n,V_n,X_n)_n$ where $X_n$ is \reviewo{as usual} the $N$-dimensional countdown process.}

	\review{Let us assume that the state of the chain $Z_n$ is known. We aim at deriving the transition to the following state $Z_{n+1}$. Let us denote by $i_n$ the index \reviewo{of the smallest coordinate} of $X_n$. It is the index of the neuron that will fire first. This neuron will fire at time $t_{n+1}=t_n+X_n^{(i_n)}$ and its countdown value $X_{n+1}^{(i_n)}$ can be computed by drawing in the law of the reset random variable. $V_{n+1}^{(i_n)}$ is simply equal to $V_r^{(i_n)}$. Let us now consider the effect of the spike fired by $i_n$ on one of its neighbors, say $j$. If the interaction is inhibitory, we can apply the results of section \ref{section:inhib} to compute the interaction random variable. If the interaction is excitatory, thanks to formula \eqref{eq:TotalProba}, we know that provided that we are able to define $V_{n+1}^{(j)}$, we can derive the interaction random variable.}

\reviewJ{The law of $V_{n+1}^{(j)}$ is uniquely determined by the value of $V_n^{(j)}$ at time $t_n$ and the value of the first passage time after $t_n$, namely $t_n+X_n^{(j)}$, because of formula \eqref{eq:Bayes}, and therefore only depend on quantities given by the state $Z_n$. The knowledge of $V_{n+1}^{(j)}$ eventually allows to compute the probability density of the next spike time of $j$ by drawing in the law of the first hitting time of the underlying membrane potential process to the spiking threshold, allowing eventually to compute $X_{n+1}^{(j)}$. }

\reviewJ{From our knowledge of the multidimensional variable $Z_n$, we have exhibited a way to draw the next state $Z_{n+1}$ of the chain, regardless of the values of $Z_k$ for $k=0,\ldots,n-1$, implying that the sequence $(Z_k)_k$ enjoys the Markov property.}
\end{proof}

\subsection{\reviewJ{LIF with instantaneous synapses}}
\reviewJ{In the case of the leaky integrate-and-fire neuron with instantaneous synaptic integration, a similar analysis leads to the fact that the new spike time after interaction has the law of the first hitting time of the membrane potential process to reach the spiking threshold $\theta$ conditionally on the fact that it has the value $V^{(j)}(t^*)+w_{ij}$ at time $t^*$. Here again, Bayes' rule allows computing in a simple form the distribution value of $V^{(j)}(t^*)$ as a function of the law of the first hitting time of the membrane potential process. Based on this law, it is very simple to extend the proof of proposition~\ref{prop:ExcitationPIF} to the present case and therefore provide a Markovian description of the spike times for this kind of network models.} 

\subsection{\reviewJ{Exponentially decaying synaptic interactions}}
\reviewJ{In the case of exponentially decaying synaptic conductances, the same framework allows deriving the interaction random variable in the excitatory case. Let us consider the case of the perfect integrate-and-fire neuron with synaptic integration and excitatory synapses. In that case, the interaction random variable is given by equation \eqref{eq:TotalProbaTwoD} and involves the value of the membrane potential at the time the postsynaptic neuron receives a spike, \reviewNew{which is evaluated using Bayes' formula in the same fashion as we did for the PIF neuron in \eqref{eq:Bayes}}.} 

\reviewJ{More precisely, assume that the neuron $j$ receives an excitatory spike from neuron $i$ at time $t^*$. Provided that this value of the membrane potential $V^{(j)}(t^*)$ and of the synaptic current $I_s^{(j)}(t^*)$ are known, the next spike time fired by neuron $j$ has the law of the the first hitting time of the spiking threshold $\theta$ of the membrane potential process with initial condition $(V_j(t^*),I_j(t^*)+w_{ij})$ at time $t^*$. The law of the $(V^{(j)}(t^*), I_s^{(j)}(t^*))$ is derived from Bayes' rule in a similar way as we did for the PIF neuron (formula \eqref{eq:Bayes}). This allows defining a Markov chain containing the times of the spikes, in a very similar fashion as done in proposition \ref{prop:ExcitationPIF}.}

\reviewJ{The case of the  leaky integrate-and-fire neuron with exponentially decaying synaptic currents can be analyzed in the same manner. }

\section*{Conclusion}
 We have developed an event-based mathematical framework for the study of stochastic integrate-and-fire neural networks. We show that in the case of the linear models treated, the spike times can be defined through a Markov chain, a method which provides an elegant way to describe the spike times and an efficient way of simulating the network. 

The approach is quite versatile, since it can deal with a large class of neuron models and different types of interactions, i.e., it can cope with transmission delays, absolute and relative refractory periods. However, it is necessary to assume that the neuron membrane has a linear free dynamics in order to be able to find the transition probabilites in closed form, as a function of \reviewNew{the first passage-time of the underlying membrane potential process to the spking threshold, which is related to the law of the interspike interval of an isolated neuron}. These variables as always well defined, in most of the cases treated, either closed form formulae or numerical methods are available, allowing efficient event-based simulation. \reviewNew{This is for instance the case of the PIF neuron with constant current, where a closed-form formula for the first passage time to a constant boundary is known.}

\review{ One of the strengths of the approach is that is makes use of the probabilistic nature of the firings. More precisely, attempts to map the actual spike times of the voltage process to a realization of a Markov chain governing the times of the spikes are out of reach. As noted in the excitatory case where an explicit dependency on the value of the membrane potential process appears, the actual spike time for a given realization of the voltage depends on the specific sample path produced. However, by releasing the constraint of pathwise agreement, we proposed a well-defined framework based on the notion of equality in law which is the correct notion of equality in a probabilistic framework. In other words, instead of mapping all the sample paths of the voltage values to one sample path of the countdown process, our construction builds a well-defined Markov chain which exactly reproduces the distribution of the spike times of the original network model. The prior to this approach is highly relevant to the underlying biological phenomenon: it consists in considering that it is the probability distribution of the spike times that contains the information of the network.}

We were able to produce expressions for the transition matrix of the spike time variables in some specific cases, something which had not been done before, and were able to address most of the cases that were addressed in the deterministic case. However, all our models remain somewhat over-simplified. The extension to a wider class of neuron models including more realistic features, in particular nonlinearities and conductance-based interactions would be highly desirable. As shown in appendix \ref{append:FurtherModels}, the extension to conductance-based noise or interactions is a very complex task. The extension to nonlinear neuron models is also difficult. In deterministic networks, the only exception is the work of Tonnelier and collaborators~\cite{tonnelier-etal:07} with quadratic neurons making use of the invertibility of the solutions, known in closed form. In the stochastic case, there is no closed-form solution to the membrane potential (now stochastic) equation which prevents from using similar methods. But an even deeper and more fundamental problem is the fact that the reset random variable will involve first hitting times of a quadratic process of type:
\[dX_t=X_t^2\,dt+\sigma \, dW_t.\]
The problem of characterizing such hitting times is still an active area of research in mathematics and no solution is currently in view. Moreover, as we mentioned above, the knowledge of the first spike time if nothing occurs meanwhile still implies that one has to cope with a complex correlation structure that makes the law of the interaction random variable quite complicated.


\review{The framework we have developed is only concerned with exact simulation of the spike times in a stochastic network. The exact simulation has been proved possible in few simple cases, mostly the same as in the deterministic case. However, we observed that our take on this problem did not allow generalization to more complex neuron models. This observation suggests to relax the constraint of exact simulation and develop approximate simulation frameworks based on spike times. The development of approximate event-based description with a controlled error on the spike times is highly desirable and the difficulties encountered in our study when dealing with exact simulation suggest further exploration along these lines. One of the possible ways to approach this problem would be to approximate the initial dynamics by one of the models that can be treated exactly, and for this purpose to find regimes and/or scales in which neurons can be approximated by linear integrate-and-fire models.The main difficulty is to deal with the complex correlation structure of the different random variables involved in this putative event-based framework which can be the source of an accumulation of  errors.} 

The Markovian model developed in this article gives a compact and simple way for describing the complex dynamics spiking networks. Besides these advantages, it is close to some renewal processes with interaction that have been studied in queueing theory. Though the results obtained in this domain do not apply so far, extensions of these theories might be reachable to prove properties such as the ergodicity of the models. Indeed, the theory developed only accounts for constant interaction random variable, i.e. whose law does not depend on the state of the network, and that have a finite expectation. This is also a very active field of research (see e.g. \cite{touboul:08b}). Our opinion is that this framework allows to use the powerful tools of communication networks theory to analyze the properties of several important models of neuronal networks in a potentially mathematically more tractable way than the study of the stochastic membrane potential. This would also open the door to the mathematical study of the macroscopic behavior of large networks using the hydrodynamics limits developed for large queuing processes, to infer and model collective behaviors of such networks (see e.g. \cite{gromoll-robert-etal:08}). The simulation efficiency of the algorithm described here opens the way to very large scale simulations of networks, a current active area of research (see e.g. \cite{izhikevich-edelman:08}), and would  allow a numerical study of finite-size effects in this type of networks.

 \section*{Acknowledgements}
 The authors warmly acknowledge Romain Brette for very insightful discussions on the concepts, Philippe Robert for interesting discussions and for reading suggestions, Olivier Rochel for his introduction to MVA Spike and for sharing his code, and Renaud Keriven and Alexandre Chariot for developing a GPU simulation code (not presented here).  \reviewo{This work was partially supported by the ERC advanced grant NerVi number 227747.}
 
\appendix
\section{Inhibitory Networks for a wider class of integrate-and-fire models}\label{append:FurtherModels}
We study more refined descriptions of the neuronal activity and show that in these cases an event-based description of the network activity may fail. 
\subsection{LIF model with general post-synaptic current pulse}
We consider a LIF neuron described by \eqref{eq:BaseLIF}, but consider that the spike integration is governed by a general postsynaptic potential function $\alpha(\cdot)$, the Green function of the differential operator $\Lop$. In that case, the synaptic input to the neuron $j$, denoted $I_a^{(j)}$, is solution of $\Lop I_a^{(j)} =0$ between two spikes, and when neuron $i$ fires a spike, the synaptic current $I_a$ is added the synaptic weight $w_{ij}$. 
In the case of the LIF neuron, the equations of the model read:
\[\begin{cases}
	\tau_i dV_t^{(i)} &= (-V_t+I_e(t)+I_a^{(i)}(t))\;dt + \sigma\;dW^{(i)}_t\\
	\Lop I_a^{(i)} (t) &=0
\end{cases}\]
When the membrane potential $V_t^{(i)}$ reaches the threshold $\theta$, a spike is fired, the membrane potential is reset to the value $V_r^{(i)}$ and the synaptic current $I_a^{(i)}$ is added the synaptic weight $w_{ij}$. The reset random variable is therefore simply the first hitting time of the Ornstein-Uhlenbeck process to the threshold, with synaptic current $I_a(t)=0$. If neuron $j$ receives a spike from neuron $i$ at time $t^*$, then the synaptic current $I_a$ is added $w_{ij}$. \review{Therefore, the updated membrane potential at time $t^*+X^{(j)}(t^*)$ reads:
\begin{equation}\label{eq:NewThet}
	\theta(t^*+X^{(j)}(t^*)) + w_{ij} e^{-X^{(j)}(t^*)}\int_{0}^{X^{(j)}(t^*)}\alpha(s)e^{s/\tau_j}\, ds
\end{equation}}
and $I_a^{(j)}(t^*+X^{(j)}(t^*))$ is the value at time $t^*+X^{(j)}(t^*)$ of the solution of the ordinary differential equation $\Lop I_a^{(j)}=0$ with the initial condition at time $t^*$ equal to $I_a^{(j)}(t^{*-})+w_{ij}$. Note that if the operator $\Lop$ is of order $d$ greater than one, the initial condition involve the derivatives of $I_a^{(j)}$ up to order $d-1$. These are unchanged by the spike reception. Hence the additional time induced by the reception of a presynaptic spike from neuron $i$ has the law of the first hitting time \review{to the boundary $\theta$ of the stochastic process $V^{(j)}$ with initial condition at time $t^*+X^{(j)}(t^*)$ given by \eqref{eq:NewThet} and $I_a^{(j)}(t^*+X^{(j)}(t^*))+w_{ij}\Psi^{j}(X^{(j)}(t^*))$ where $\Psi$ denotes the flow of the linear equation $\Lop \Psi = 0$}. This random variable has a law independent of the value of the membrane potential at time $t^*$ and therefore one can build a Markov chain governing the times of the spikes. In details, in order to take into account this synaptic integration of spikes in our framework, we have to extend the phase space of our Markov chain. The Markovian variable we consider is the process $(X_t, I_a(t))_{t\geq 0}$. When a neuron $i$ elicits a spike, i.e. when its countdown reaches $0$ at time $t^*$, its countdown value is reset by drawing from the law of the first hitting time of the membrane potential with initial condition $(V_r^{(i)}, I_a^{(i)}(t^*))$ to the threshold and for all neuron $j \in \V(i)$, their spike-induced current $I^{(j)}_a(t^*)$ are instantaneously updated by adding the synaptic efficiency $w_{ij}$ : $I_a^{(j)}(t^*) = I_a^{(j)}(t^{*-}) + w_{ij}$. Simulating this Markov process, that can be sampled at the times of the spike emission, is equivalent from the spikes point of view as simulating the whole membrane potential process (see theorem \ref{thm:FundamentalEquivalence}).

\subsection{LIF models with noisy conductances}

We consider the case of noisy conductance-based models. In this case, the membrane potential is solution of equation:
\begin{equation}\label{eq:NoisyConductances}
\left \{
\begin{array}{lll}
dV^{(i)}_t &=& (I_e^{(i)}(t) - \lambda_i (V^{(i)}_t-V_{rev}^{(i)}))\, dt + I_s^{(i)}(t)\,dt + \sigma_i \, g_i \, (V^{(i)}_t-V_{rev}^{(i)}) \, dW_t^i \\
V^{(i)}(t^-) &=& \theta \Rightarrow V^{(i)}(t) = V_r^{(i)} 
\end{array}
\right. ,
\end{equation}
The term $I_s^{(i)}$ corresponds to the current generated by the reception of spikes emitted from neurons in the network. $V_{rev}^{(i)}$ is the reversal potential of the synapse.

Between two spikes, the stochastic differential equation can be integrated in closed-form. For instance, provided that  $V^{(i)}_{t_0}$ is known, we have for $t\geq t_0$ (see e.g. \cite{karatzas-shreve:87}): 
\[V^{(i)}_t = Z^{(i)}_{t-t_0} \Big( V^{(i)}_{t_0} +\int_{t_0}^t \frac 1 {Z_{u-t_0}} (I_e(u)+I_s(u))\,du \Big)\]
where $Z_t^{(i)}$ is the process:
\[Z_t^{(i)} = \exp \Big(-(\lambda_i+\frac{g_i^2\sigma_i^2}{2}) \; t + g_i\,\sigma_i W^{(i)}_t\Big).\]
This models makes it natural to consider that the spike reception modifies the conductance of the network. 
This synapse model is considered first, and instantaneous current synapses next. 

\subsubsection{Conductance-based synapses}
When neuron $j$ receives a spike from one of its neighbors $i$, a current is generated, which has the value $w_{ij} g (V^{(j)} - V_{rev})$ . Note that we artificially introduced $V_{rev}$ in the leak term, which amounts to formally changing the current $I_e^{(i)}$, in order to integrate the equation \review{in a simpler way}. When neuron $j$ receives a spike at time $t^*$ from neuron $i$, the conductance is hence increased by a coefficient $w_{ij}\,g$. The solution of the membrane potential equation after time $t^*$ reads:
\[
V^{(j)}(t+t^*) = V_{(j)}^* Z_t^{(j)} + \int_0^t I_e^{(j)}(s+t^*)Z_{t-s}^{(j)}\,ds
\]
where $Z_t^{(j)} = \exp \{ -(\lambda_j+ \frac{\sigma_j^2}{2} -w_{ij} g) (t-t^*) + \sigma_j W_t^{(j)}\}$. The reception of the spike therefore modifies the value of the membrane potential, which would have been equal to
\[
\widetilde{V}^{(j)}(t+t^*) = V_{(j)}^* \, {Z_t}e^{w_{ij}\,g\,t} + \int_0^t I_e^{(j)}(s+t^*)\,{Z}_{t-s}e^{w_{ij}\,g\,(t-s)}\,ds
\]
if neuron $j$ did not receive any spike. At time $X^{(j)}_*$, the value of $\widetilde{V}^{(j)}(X^{(j)}_*+t^*)$ is equal to $\theta(X^{(j)}_*+t^*)$, but after inhibition at time $t^*$, the actual value of the membrane potential reads:
\[V^{(j)}(X^{(j)}_*+t^*) = \theta e^{w_{ij}\,g \,X^{(j)}_*} + \int_{0}^{X^{(j)}_*}I_e^{(j)}(s+t^*)Z_{t-s}(e^{w_{ij}\,g\,s}-1)\,ds \]
This value therefore depends on the history of the Brownian motion from time $t^*$ on. We are interested in finding the first spike time if nothing occurs in the network meanwhile. This is described by the random variable:
\begin{align*}
	\eta_{ij}& =\inf\{t>t^*, V^{(j)}_{t}=\theta(t) \vert \widetilde{V}^{(j)}(X^{(j)}_{t^*}+t^*)=\theta(X^{(j)}_{t^*}+t^*)\}\\
	&= X^{(j)}_{t^*} + \inf\{t>t^*, V^{(j)}_{t}=\theta(t) \vert \widetilde{V}^{(j)}(X^{(j)}_{t^*}+t^*)=\theta(X^{(j)}_{t^*}+t^*)\}\\
\end{align*}
This case is therefore substantially more complex than the previous ones. Indeed, conditioning on the event that $\widetilde{V}^{(j)}(t^*+X^{(j)}_{t^*})$ is the first crossing time of the threshold of $V^{(j)}$ conditions the trajectories of the Brownian motion in the interval $[t^*, t^*+X^{(j)}_{t^*}]$, and we observe that the value of the updated process $V^{(j)}$ depends on the whole history of the Brownian motion in this interval (except in the very particular case where $I_e=0$, in which case the problem can be treated as previously). The problem of finding the law of the interaction random variable is therefore very tricky because of this complicated conditioning. 

\subsubsection{{Noisy Conductances and current-based synaptic interactions}}
We now consider the case of instantaneous interactions. In that case, if neuron $j$ receives a spike from neuron $i$ at time $t^*$, its membrane potential is instantaneously increased by a quantity equal to $w_{ij}$. Similarly to the computations done in the previous case, the updated membrane potential now reads:
\[V^{(j)}(t+t^*) = (V_{(j)}^*+w_{ij}) Z_t^{(j)} + \int_0^t I_e^{(j)}(s+t^*)Z_{t-s}^{(j)}\,ds\] 
and therefore at time $X^{(j)}_{t^*}$, this value reads:
\[V^{(j)}(X^{(j)}_{t^*}+t^*) = (\theta + w_{ij}) Z_{X^{(j)}_{t^*}}^{(j)}\] 
We therefore need to characterize the probability distribution of $Z_{X^{(j)}_{t^*}}^{(j)}$ conditioned on the fact that the neuron elicited a spike at time $X^{(j)}_{t^*}$ if it did not receive a spike at time $t^*$ from neuron $i$. Here again, this problem depends in an intricate fashion on the distribution of the values of the underlying Brownian motion $W^{(j)}$ at this time conditioned on the fact that the first crossing time of the process with the threshold was $X^{(j)}_{t^*}$. To the best of our knowledge, no solution is available so far.
\subsection{Conclusion}
The current-based synapses are a necessary hypothesis in order to be able to express in closed-form the random variables involved in the transitions of the countdown process. The same problem arises in the case of nonlinear models: it remains very difficult to express the interaction random variable conditioned on the value of the countdown process at the time of the spike reception. 

It is important to note here that this limitation is also present in the deterministic case. Extensions of the framework of simple linear integrate-and-fire models are scarce and apply to very particular models. The only successful extension to our knowledge was done by Tonnelier and collaborators \cite{tonnelier-etal:07} in the case of the quadratic integrate-and-fire neuron with constant input and constant threshold. It is based on the fact that the authors provide a closed-form expression for the membrane-potential. This fact is no more possible in the stochastic case, because even if we were to find a closed form expression for the membrane potential, the integration random variable will be very complex to express because of the conditioning on the value of the countdown variable as discussed previously.


\section{Including Synaptic Delays and the Refractory Period}\label{append:SynDelays}
In this appendix we include two biologically plausible phenomena in the description of the network activity, and see how this affects the Markovian framework. 

\subsection{Refractory period}
The refractory period is a transient phase just after the firing during which it is either impossible or difficult to \review{communicate} with the cell. This phenomenon is linked with the dynamics of ion channels and the hyperpolarization phase of the spike emission, lasts a few milliseconds, and prevents the neuron from firing spikes at an arbitrary high firing rate. It can be decomposed into two phases: the \emph{absolute} refractory period, which is a constant period of time corresponding loosely to the hyperpolarization of the neuron during which is it impossible to excite or inhibit the cell no matter how great the stimulating current applied is (see for instance  \cite[chapter 9]{kandel-schwartz-etal:00} for a further biological discussion of the phenomenon and \cite{gerstner-kistler:02b, arbib:98} for a discussion on the modeling this refractory period). Immediately after this phase begins the \emph{relative} refractory period during which the initiation of a second action potential is inhibited but still possible. Modeling this relative refractory period amounts to considering that the synaptic inputs received at the level of the cell are weighted by a function depending on the time elapsed since the spike emission. 

For technical reasons, the relative refractory period is taken into account only for the spike integration, and not for the noise integration. This assumption does not affect significantly the dynamics of the network, since the probability that the noisy current integrated during a time period as short as $1$ or $2$ms be substantial is very small. For the spike integration this remark is no more valid, since a single spike induces large changes in the membrane voltage. During the relative refractory period, we consider that synaptic efficiencies are weighted by a function depending on the time elapsed since the last spike was fired. We denote this function $\kappa(t)$, following the notation of Gerstner and Kistler in \cite{gerstner-kistler:02b}. In our case this function is unspecified, it is zero at $t=0$ and increases to $1$ with a characteristic time of around $2ms$. It can be of bounded support or defined on $\R$ (see figure \ref{fig:kappa}).
\MyFigure{0.4}{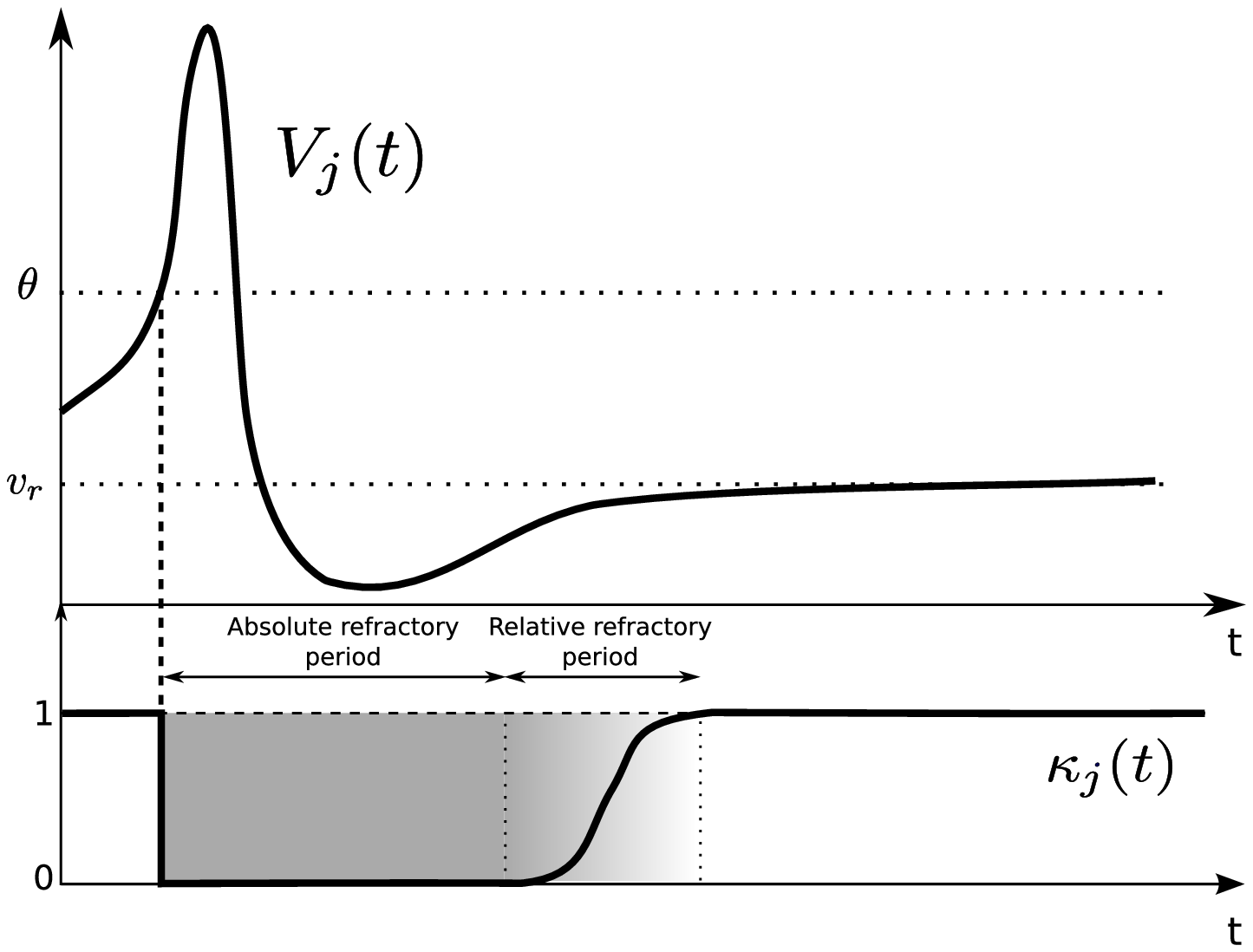}{Refractory period}{The refractory period at a spike emission, and the related $\kappa$ function weighting the synaptic inputs}{fig:kappa}
%
%
We will see how taking in account this effect modifies the above instantaneous analysis, after introducing another essential effect occurring at the same time scale as the refractory period: synaptic delays.

\subsection{Synaptic delays}
\begin{figure}
 \begin{center}
 \includegraphics[width=.4\textwidth]{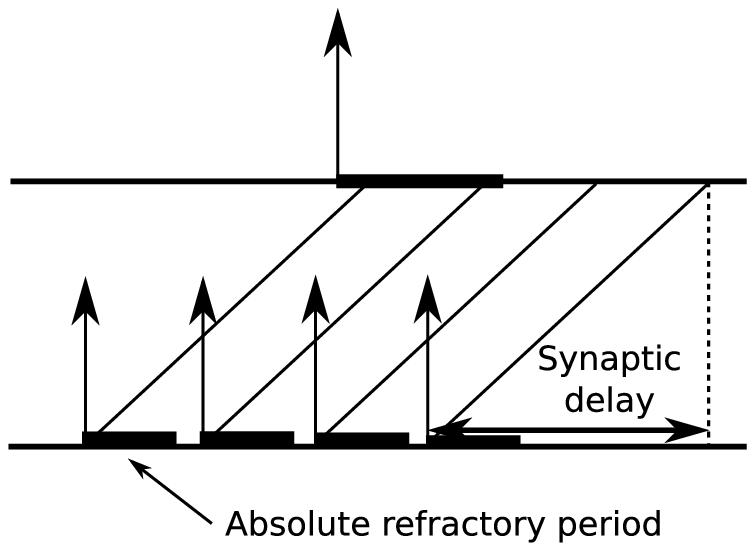}
 \end{center}
 \caption{\john{There is a finite number of incoming spikes emitted before the postsynaptic cell fires.}}
 \label{fig:SpikesTraversent}
\end{figure}

Delays are known to be very important, for instance in shaping spatio-temporal dynamics of neuronal activity \cite{roxin-brunel-etal:05} or for generating global oscillations \cite{brunel-hakim:99}. These synaptic delays affect the interaction variable in a quite intricate fashion
%
%
by adding a non-trivial memory-like phenomenon in the network. As a consequence of this we are not able to update instantaneously the countdown variable at each spike time: the transmission delays imply that one needs to keep the memory of a certain number of spikes times. Fortunately, because of the absolute refractory period, we only have to take into account a finite number of spikes that can possibly affect the postsynaptic potential after it elicits a spike (see figure \ref{fig:SpikesTraversent}).
The maximal number of spikes concerned is given by $M \eqdef \john{\max_{i,j}} \lfloor \frac{\Delta_{ij}}{R_j} \rfloor$ where $R_j$ is  the length of the absolute refractory period and $\lfloor x \rfloor$ is the floor function, i.e. the largest integer smaller than or equal to $x$. 

Instead of considering the last firing times variables which contained the last firing time for each neuron of the network, we now consider the last $M$ firing times variables. To this end, we define the matrix $H_M \in \R^{N \times M}$ whose row $i$ contains the $M$ last firing times of neuron $i$. At the initial time, the $M$ components of this row are set to the value $ \min_{ij} \{ -R_i -\Delta_{ij} \}$. If the neuron spikes at time $t_i$, then each component of the row are modified: for all $k \in \{2, \ldots, N\}$, $H_{i,k-1}=H_{i,k}$ and $H_{i,M} = t_i$, and between two consecutive spikes of neuron $i$, the elements of the row $i$ remain constant. 

Provided that the spike times are known, this matrix is  a Markov chain.

Indeed, let us denote by $X^n$ the countdown process of the network. The updates of the Markov chain \john{$(X,H_M)$} occur at spike emission times and spike  arrivals at postsynaptic cells. The next spike if no delayed interaction occurs will be fired after a time given by $\tau = \min_{i} X_i^n$, and the first arrival of a possible spike at a cell is given by 
\[\nu = \min_{\substack{i,j \in \{1,\,\ldots\,,N\}\\ k\in \{1,\,\ldots,\,M\}}} \{ x= H_{i,k} + \Delta_{ij} - t; x>0\}\]
\begin{itemize}
\item
If this set is empty, the $\min$ is set to $+\infty$. If $\tau < \nu$, a spike is fired by the neuron $i$ having the lowest countdown value. At this instant, the related countdown variable of neuron $i$ is instantaneously reset by drawing in the law of the reset variable as described in section \ref{ssect:Reset}, and row $i$ of the matrix $H$ is updated as indicated above. Other variables, such as the absolute time, are also to be updated at this point (e.g.  the absolute time is updated to $t_{m-1}+X_i$, \ldots).

\item
If $\nu<\tau$, assume that the minimum is achieved for the value $H_{i,k}+\Delta_{ij}$ for some $i,j,k$. This means that the $k^{th}$ latest spike of neuron $i$ reaches the cell $j$ and affects the cell in the same fashion as if there were no delay and a spike was emitted at this time. Therefore, the related interaction variable \john{$\eta_{ij}$} of this connection will be added, and the countdown value of neuron $j$ is updated together with the other Markov chain variables. The time is advanced to $t+\nu$. Note also that the min can be achieved at many different 3-tuples $(i,j,k)$ at the same time. Moreover, it is also possible that an excitatory interaction makes a postsynaptic neuron fire instantaneously at the reception of the spike. All these cases can be treated sequentially, by iterating the mechanism we just described. Nevertheless, we are ensured that no avalanche can occur, because of the absolute refractory period and of the delays.
\end{itemize}

\subsection{The reset random variable}
Let us consider the effect of these features from the countdown process viewpoint. The reset variable is only affected by the absolute refractory period, and in a very simple way. Indeed, we formally consider that the neuron $i$ is stuck at its reset value $V_r^{(i)}$ during a fixed period of time $R_i$ after having fired. After this period, the neuron membrane potential follows the evolution that depends on the particular model chosen. Therefore, the time of the next spike starting from time $t+R_i$ has the same law as the reset variable in the case where we do not take into account the refractory period and the synaptic delay, i.e. it has the law of the first hitting time, noted $\tau_i$, of the membrane potential process to the spike threshold with the time-shifted input $I_e(t+t^*+R_i)$. The new reset variable of the related countdown process has simply the law of $Y_i = \tau_i+R_i$.

The initialization random variable is clearly unchanged by taking into account synaptic delays, refractory periods and excitatory interactions. 

However, the case of the interaction variable is a little bit more intricate, as shown below.

\subsection{Interaction random variable}
Let us consider the effect of a spike emitted by neuron $i$ and reaching cell $j$ at time $t$ in an inhibitory network. This spike will affect the membrane potential depending on the connectivity model chosen but corresponding to a synaptic weight $\kappa_j(t-t_j)w_{ij}$ where $t_j$ is the time of the last spike emitted by the cell $j$. 

Since we assumed the connection inhibitory, then the interaction random variable is readily deduced from the analysis of section \ref{section:inhib} and of appendix \ref{append:FurtherModels}, by changing the synaptic weight value $w_{ij}$ to $w_{ij}\kappa_j(t_i-t_j)$. Therefore in addition to the complementary variables necessary to define the countdown value, we need to keep in memory the last spike time for each neuron to define the synaptic weights and therefore the interaction random variable. Note that this random variable is almost surely equal to zero, whatever the neuron model considered, if the spike arrives during the absolute refractory period (i.e. $t\in [t_j, t_j+R_j]$), since in that case $\kappa(t-t_j)=0$.

The spike transmission from the presynaptic cell to the postsynaptic one depends on the distance between the two cells, the speed of transmission of the signal along the axon and the transmission time at the synapse, and has a typical duration of a few milliseconds. 
To model the synaptic delay we consider that spikes emitted by a neuron affect the postsynaptic neurons after some delay $\Delta_{ij}$ which can depend on both the presynaptic and the postsynaptic neurons.

If neuron $i$ fires a spike at time $t_i$, its effect on the postsynaptic neuron $j$ depends on the synaptic delay $\Delta_{ij}$, the countdown value of neuron $j$ at this time $X^{(j)}(t_i)$, and the time of the last spike emitted by $j$:
\begin{enumerate}
 \item If $\Delta_{ij} < X^{(j)}(t)$, then the reception of a spike at time \john{$t_i$} acts on the post-synaptic neuron at time $t+\Delta_{ij}$ in the same way as discussed for the different models considered in section \ref{section:theoretical}, but the effect can now be either excitatory or inhibitory, with a synaptic efficiency $w_{ij}\kappa_j(t_i+\Delta_{ij}-t_j)$. 
 \item If $\Delta_{ij} > X^{(j)}(t_i)$, the postsynaptic neuron will fire before receiving the spike from the presynaptic cell $i$, and it will act on the postsynaptic cell membrane with an efficiency $w_{ij}\kappa_j(t_i+\Delta_{ij} - X_j)$ at time $t_i+\Delta_{ij}$. 
\end{enumerate}


 
\bibliographystyle{apa}
%

\begin{thebibliography}{}

\bibitem[\protect\astroncite{Arbib}{1998}]{arbib:98}
Arbib (1998).
\newblock {\em The Handbook of Brain Theory and Neural Networks}.
\newblock MIT Press.

\bibitem[\protect\astroncite{Asmussen and Turova}{1998}]{asmussen-turova:98}
Asmussen, S. and Turova, T.~S. (1998).
\newblock Stationarity properties of neural networks.
\newblock {\em Journal of Applied Probabilities}, 35:783--794.

%


\bibitem[\protect\astroncite{Billingsley}{1999}]{billingsley:99}
Billingsley, P. (1999).
\newblock {\em Convergence of Probability Measures}.
\newblock Wiley series in probability and statistics.

\bibitem[\protect\astroncite{Brette}{2006}]{brette:06}
Brette, R. (2006).
\newblock Exact simulation of integrate-and-fire models with synaptic
  conductances.
\newblock {\em Neural Computation}, 18(8):2004--2027.

\bibitem[\protect\astroncite{Brette}{2007}]{brette:07}
Brette, R. (2007).
\newblock Exact simulation of integrate-and-fire models with exponential
  currents.
\newblock {\em Neural Computation}, 19(10):2604--2609.

\bibitem[\protect\astroncite{Brette and Gerstner}{2005}]{brette-gerstner:05}
Brette, R. and Gerstner, W. (2005).
\newblock Adaptive exponential integrate-and-fire model as an effective
  description of neuronal activity.
\newblock {\em Journal of Neurophysiology}, 94:3637--3642.

\bibitem[\protect\astroncite{Brette et~al.}{2007}]{brette-rudolph-etal:07}
Brette, R., Rudolph, M., Carnevale, T., Hines, M., Beeman, D., Bower, J.~M.,
  Diesmann, M., Morrison, A., Goodman, P.~H., Jr., F. C.~H., Zirpe, M.,
  Natschl{\"a}ger, T., Pecevski, D., Ermentrout, B., Djurfeldt, M., Lansner,
  A., Rochel, O., Vieville, T., Muller, E., Davison, A.~P., Boustani, S.~E.,
  and Destexhe, A. (2007).
\newblock Simulation of networks of spiking neurons: a review of tools and
  strategies.
\newblock {\em Journal of Computational Neuroscience}, 23(3):349--398.

\bibitem[\protect\astroncite{Brunel and Hakim}{1999}]{brunel-hakim:99}
Brunel, N. and Hakim, V. (1999).
\newblock Fast global oscillations in networks of integrate-and-fire neurons
  with low firing rates.
\newblock {\em Neural Computation}, 11:1621--1671.

\bibitem[\protect\astroncite{Brunel and Sergi}{1998}]{brunel-sergi:98}
Brunel, N. and Sergi, S. (1998).
\newblock Firing frequency of leaky integrate and fire neurons with synaptic
  current dynamics.
\newblock {\em J. Theor. Biol.}, 195(87--95).

\bibitem[\protect\astroncite{Burkitt}{2006}]{burkitt:06}
Burkitt, A. (2006).
\newblock A review of the integrate-and-fire neuron model: I. homogeneous
  synaptic input.
\newblock {\em Biological Cybernetics}, 95(1):1--19.

\bibitem[\protect\astroncite{Chariot et~al.}{2006}]{chariot-keriven-etal:06}
Chariot, A., Keriven, R., and Brette, R. (2006).
\newblock Simulation rapide de mod{\`e}les de neurones impulsionnels sur carte
  graphique.

\bibitem[\protect\astroncite{Claverol et~al.}{2002}]{claverol-etal:02}
Claverol, E., Brown, A., and Chad, J. (2002).
\newblock Discrete simulation of large aggregates of neurons.
\newblock {\em Neurocomputing}, 47:277--297.

\bibitem[\protect\astroncite{Cottrell}{1992}]{cottrell:92}
Cottrell, M. (1992).
\newblock Mathematical analysis of a neural network with inhibitory coupling.
\newblock {\em Stochastic Processes and their Applications}, 40:103--127.

\bibitem[\protect\astroncite{Cottrell and Turova}{2000}]{cottrell-turova:00}
Cottrell, M. and Turova, T. (2000).
\newblock Use of an hourglass model in neuronal coding.
\newblock {\em Journal of applied probability}, 37:168--186.

\bibitem[\protect\astroncite{Davis}{1984}]{davis:84}
Davis, M. (1984).
\newblock Piecewise-deterministic markov processes: A general class of
  non-diffusion stochastic models.
\newblock {\em Journal of the Royal Society, Series B (Methodological)},
  46:353--388.

\bibitem[\protect\astroncite{Delorme and Thorpe}{2001}]{delorme-thorpe:01}
Delorme, A. and Thorpe, S. (2001).
\newblock Face processing using one spike per neuron: resistance to image
  degradation.
\newblock {\em Neural Networks}, 14:795--804.

\bibitem[\protect\astroncite{Delorme and Thorpe}{2003}]{delorme-thorpe:03}
Delorme, A. and Thorpe, S. (2003).
\newblock 57 spikenet: an event-driven simulation package for modelling large
  networks of spiking neurons.
\newblock {\em Network}, 14(4):613--627.

\bibitem[\protect\astroncite{Destexhe et~al.}{}]{destexhe-mainen-etal:98}
Destexhe, A., Mainen, Z.~F., and Sejnowski, T.~J.
\newblock {\em Methods in Neuronal Modeling}, chapter Kinetic models of
  synaptic transmission, pages 1--25.

\bibitem[\protect\astroncite{Fabre-Thorpe
  et~al.}{1998}]{fabre-thorpe-richard-etal:98}
Fabre-Thorpe, M., Richard, G., and Thorpe, S. (1998).
\newblock Rapid categorization of natural images by rhesus monkeys.
\newblock {\em Neuroreport}, 9(2):303--308.

\bibitem[\protect\astroncite{Fricker et~al.}{1994}]{fricker-robert-etal:94}
Fricker, C., Robert, P., Saada, E., and Tibi, D. (1994).
\newblock Analysis of some networks with interaction.
\newblock {\em Annals of Applied Probability}, 4:1112--1128.

\bibitem[\protect\astroncite{Gerstner and
  Kistler}{2002a}]{gerstner-kistler:02b}
Gerstner, W. and Kistler, W. (2002a).
\newblock {\em Spiking Neuron Models}.
\newblock Cambridge University Press.

\bibitem[\protect\astroncite{Gerstner and Kistler}{2002b}]{gerstner-kistler:02}
Gerstner, W. and Kistler, W.~M. (2002b).
\newblock Mathematical formulations of hebbian learning.
\newblock {\em Biological Cybernetics}, 87:404--415.

\bibitem[\protect\astroncite{Gobet}{2000}]{gobet:00}
Gobet, E. (2000).
\newblock {Weak approximation of killed diffusion using Euler schemes}.
\newblock {\em Stochastic Processes and their Applications}, 87(2):167--197.

\bibitem[\protect\astroncite{Goldman}{1971}]{goldman:71}
Goldman, M. (1971).
\newblock On the first passage of the integrated {W}iener process.
\newblock {\em Ann. Mat. Statist.}, 42:2150--2155.

\bibitem[\protect\astroncite{Goodman and Brette}{2008}]{goodman-brette:08}
Goodman, D. and Brette, R. (2008).
\newblock Brian: a simulator for spiking neural networks in python.
\newblock {\em Frontiers in Neuroinformatics (in preparation)}.

\bibitem[\protect\astroncite{Gromoll et~al.}{2008}]{gromoll-robert-etal:08}
Gromoll, H., Robert, P., and Zwart, B. (2008).
\newblock Fluid limits for processor sharing queues with impatience.
\newblock {\em Mathematics of Operations Research}, 33(2):375--402.

\bibitem[\protect\astroncite{Holden}{1976}]{holden:76}
Holden, A. (1976).
\newblock {Models of the Stochastic Activity of Neurones}.
\newblock {\em Lecture Notes in Biomathematics}, 12:1--368.

\bibitem[\protect\astroncite{Izhikevich}{2003}]{izhikevich:03}
Izhikevich, E. (2003).
\newblock Simple model of spiking neurons.
\newblock {\em IEEE Transactions on Neural Networks}, 14(6):1569--1572.

\bibitem[\protect\astroncite{Izhikevich and
  Edelman}{2008}]{izhikevich-edelman:08}
Izhikevich, E.~M. and Edelman, G.~M. (2008).
\newblock {L}arge-scale model of mammalian thalamocortical systems.
\newblock {\em Proc Natl Acad Sci U S A}, 105(9):3593--3598.

\bibitem[\protect\astroncite{Kandel et~al.}{2000}]{kandel-schwartz-etal:00}
Kandel, E., Schwartz, J., and Jessel, T. (2000).
\newblock {\em Principles of Neural Science}.
\newblock McGraw-Hill, 4th edition.

\bibitem[\protect\astroncite{Karatzas and Shreve}{1987}]{karatzas-shreve:87}
Karatzas, I. and Shreve, S. (1987).
\newblock {\em Brownian motion and stochatic calculus}.
\newblock Springer.

\bibitem[\protect\astroncite{Kloeden and Platen}{1992}]{kloeden-platen:92}
Kloeden, P. and Platen, E. (1992).
\newblock {\em Numerical solution of stochastic differential equations}.
\newblock Springer-Verlag.

\bibitem[\protect\astroncite{Lachal}{1991}]{lachal:91}
Lachal, A. (1991).
\newblock Sur le premier instant de passage de l'int{\'e}grale du mouvement
  brownien.
\newblock {\em Annales de l'IHP, section B}, 27:385--405.

\bibitem[\protect\astroncite{Lachal}{1996}]{lachal:96b}
Lachal, A. (1996).
\newblock {Sur la distribution de certaines fonctionnelles de l'int'egrale du
  mouvement Brownien avec d'erives parabolique et cubique}.
\newblock {\em Communications on Pure and Applied Mathematics}, 49:1299--1338.

\bibitem[\protect\astroncite{Lapicque}{1907}]{lapicque:07}
Lapicque, L. (1907).
\newblock Recherches quantitatifs sur l'excitation des nerfs traitee comme une
  polarisation.
\newblock {\em J. Physiol. Paris}, 9:620--635.


\bibitem[\protect\astroncite{Makino}{2003}]{makino:03}
Makino, T. (2003).
\newblock A discrete-event neural network simulator for general neuron models.
\newblock {\em Neural Comput. and Applic.}, 11:210--223.

\bibitem[\protect\astroncite{Marian et~al.}{2002}]{marian-etal:02}
Marian, I., Reilly, R., and Mackey, D. (2002).
\newblock Efficient event-driven simulation of spiking neural networks.
\newblock In {\em Proceedings of the 3rd WSEAS International Conference on
  Neural Networks and Applications}.

\bibitem[\protect\astroncite{McKean}{1963}]{mckean:63}
McKean, H.~P. (1963).
\newblock A winding problem for a resonator driven by a white noise.
\newblock {\em J. Math. Kyoto Univ.}, 2:227--235.

\bibitem[\protect\astroncite{Plesser}{1999}]{plesser:99}
Plesser, H.~E. (1999).
\newblock {\em Aspects of signal processing in noisy neurons}.
\newblock PhD thesis, Georg-August-Universit{\"a}t.

\bibitem[\protect\astroncite{Reutimann
  et~al.}{2003}]{reutimann-giugliano-etal:03}
Reutimann, J., Giugliano, M., and Fusi, S. (2003).
\newblock Event-driven simulation of spiking neurons with stochastic dynamics.
\newblock {\em Neural Computation}, 15:811--830.

\bibitem[\protect\astroncite{Ricciardi and Smith}{1977}]{ricciardi:77}
Ricciardi, L. and Smith, C. (1977).
\newblock {\em Diffusion processes and related topics in biology}.
\newblock Springer-Verlag New York.

\bibitem[\protect\astroncite{Rochel}{2004}]{rochel:04}
Rochel, O. (2004).
\newblock {\em Une approche \'ev\'enementielle pour la mod\'elisation et la
  simulation de neurones impulsionnels}.
\newblock PhD thesis, Universit\'e Henri Poincar\'e - Nancy 1.

\bibitem[\protect\astroncite{Rochel and Martinez}{2003}]{rochel-martinez:03}
Rochel, O. and Martinez, D. (2003).
\newblock An event-driven framework for the simulation of networks of spiking
  neurons.
\newblock In {\em Proc. 11th European Symposium on Artificial Neural Networks},
  pages 295--300.

\bibitem[\protect\astroncite{Rolls and Deco}{2010}]{rolls-deco:10}
Rolls, E. and Deco, G. (2010).
\newblock {\em The noisy brain: stochastic dynamics as a principle of brain
  function}.
\newblock Oxford university press.

\bibitem[\protect\astroncite{Roxin et~al.}{2005}]{roxin-brunel-etal:05}
Roxin, A., Brunel, N., and Hansel, D. (2005).
\newblock {Role of Delays in Shaping Spatiotemporal Dynamics of Neuronal
  Activity in Large Networks}.
\newblock {\em Physical Review Letters}, 94(23):238103.

\bibitem[\protect\astroncite{Rudolph and Destexhe}{2006}]{rudolph-destexhe:06}
Rudolph, M. and Destexhe, A. (2006).
\newblock Analytical integrate-and-fire neuron models with conductance-based
  dynamics for event-driven simulation strategies.
\newblock {\em Neural Computation}, 18:2146--2210.

\bibitem[\protect\astroncite{Shadlen and Newsome}{1994}]{shadlen-newsome:94}
Shadlen, M.~N. and Newsome, W.~T. (1994).
\newblock Noise, neural codes and cortical organization.
\newblock {\em Curr Opin Neurobiol}, 4(4):569--579.

\bibitem[\protect\astroncite{Softky and Koch}{1993}]{softky-koch:93}
Softky, W.~R. and Koch, C. (1993).
\newblock The highly irregular firing of cortical cells is inconsistent with
  temporal integration of random epsps.
\newblock {\em Journal of Neuroscience}, 13:334--350.

\bibitem[\protect\astroncite{Taillefumier}{2011}]{PersoCom}
Thibaud Taillefumier
\newblock{\em Personal Communication. }

\bibitem[\protect\astroncite{Thorpe et~al.}{2001}]{thorpe-delorme-etal:01}
Thorpe, S., Delorme, A., and VanRullen, R. (2001).
\newblock Spike based strategies for rapid processing.
\newblock {\em Neural Networks}, 14:715--726.

\bibitem[\protect\astroncite{Tonnelier et~al.}{2007}]{tonnelier-etal:07}
Tonnelier, A., Belmabrouk, H., and Martinez, D. (2007).
\newblock Event-driven simulations of nonlinear integrate-and-fire neurons.
\newblock {\em Neural Computation}, 19(12):3226--3238.

\bibitem[\protect\astroncite{Touboul}{2008a}]{touboul:08}
Touboul, J. (2008a).
\newblock Bifurcation analysis of a general class of nonlinear
  integrate-and-fire neurons.
\newblock {\em SIAM Journal on Applied Mathematics}, 68(4):1045--1079.

\bibitem[\protect\astroncite{Touboul}{2008b}]{touboul:08b}
Touboul, J. (2008b).
\newblock {\em Nonlinear and stochastic models in neuroscience}.
\newblock PhD thesis, Ecole Polytechnique.

\bibitem[\protect\astroncite{Touboul and Faugeras}{2007}]{touboul-faugeras:07b}
Touboul, J. and Faugeras, O. (2007).
\newblock The spikes trains probability distributions: a stochastic calculus
  approach.
\newblock {\em Journal of Physiology, Paris}, 101/1-3:78--98.

\bibitem[\protect\astroncite{Touboul and Faugeras}{2008}]{touboul-faugeras:08}
Touboul, J. and Faugeras, O. (2008).
\newblock First hitting time of double integral processes to curved boundaries.
\newblock {\em Advances in Applied Probability}, 40(2):501--528.

\bibitem[\protect\astroncite{Tuckwell}{1988}]{tuckwell:88}
Tuckwell, H.~C. (1988).
\newblock {\em Introduction to theoretical neurobiology}.
\newblock Cambridge University Press.

\bibitem[\protect\astroncite{Turova}{2000}]{turova:00}
Turova, T. (2000).
\newblock Neural networks through the hourglass.
\newblock {\em BioSystems}, 58:159--165.

\bibitem[\protect\astroncite{Turova}{1996}]{turova:96}
Turova, T.~S. (1996).
\newblock Analysis of a biological plausible neural network via an hourglass
  model.
\newblock {\em Markov Processes and related fields}, 2:487--510.

\bibitem[\protect\astroncite{van Rotterdam
  et~al.}{1982}]{rotterdam-lopes-da-silva-etal:82}
van Rotterdam, A., Lopes~da Silva, F., van~den Ende, J., Viergever, M., and
  Hermans, A. (1982).
\newblock A model of the spatial-temporal characteristics of the alpha rhythm.
\newblock {\em Bulletin of Mathematical Biology}, 44(2):283--305.

\bibitem[\protect\astroncite{Vi\'eville and Rochel}{2006}]{vieville-rochel:06}
Vi\'eville, T. and Rochel, O. (2006).
\newblock One step towards an abstract view of computation in spiking
  neural-networks.
\newblock In {\em International Conf. on Cognitive and Neural Systems}.

\bibitem[\protect\astroncite{Watts}{1994}]{watts:94}
Watts, L. (1994).
\newblock Event-driven simulation of networks of spiking neurons.
\newblock {\em Advances in Neural Information Processing System}, pages
  927--934.

\end{thebibliography}

\end{document}